\documentclass[10pt,twocolumn]{article} 
\usepackage{abstract}
\usepackage[utf8]{inputenc} 
\usepackage{amsxtra} 
\usepackage{amsthm}
\usepackage{amssymb}
\usepackage{amsfonts}
\usepackage{graphicx}
\usepackage{rotating}
\usepackage{color}
\usepackage{relsize}
\usepackage{epsfig}
\usepackage{subcaption} 
\usepackage{verbatim}
\usepackage[printonlyused]{acronym} 
\usepackage{setspace}
\usepackage{epstopdf}
\usepackage{authblk}
\usepackage{sectsty}
\usepackage[colorinlistoftodos]{todonotes}
\usepackage[colorlinks=true,allcolors=black]{hyperref}
\usepackage{textcomp,marvosym}
\usepackage{soul}
\usepackage{tikz}
\usetikzlibrary{shapes,arrows,automata,positioning}
\usepackage{mathtools}

\sectionfont{\fontsize{12}{15}\selectfont} 
\subsectionfont{\fontsize{10}{12}\selectfont} 

\usepackage{geometry} 
\definecolor{blue}{RGB}{0, 0, 150}
\definecolor{green}{RGB}{0,150,0}
\definecolor{red}{RGB}{200, 0, 0}
\definecolor{black}{RGB}{0, 0, 0}
     
\newcommand{\pr}[1]{\mathbb{P}\left\{#1\right\}}
\newcommand{\E}[1]{\mathbb{E}\left[#1\right]}

\renewcommand{\l}{\left}
\renewcommand{\r}{\right}
\newcommand{\subeq}[2]{\begin{subequations}\label{#2}\begin{align}#1\end{align}\end{subequations}}
\newcommand{\al}[2]{\begin{align}#1\label{#2}\end{align}}
\newcommand{\eq}[2]{\begin{equation}#1\label{#2}\end{equation}}

\newcommand{\dd}{{\rm d}}

\begin{document}

\title{Efficient analysis of stochastic gene dynamics in the non-adiabatic regime using piecewise deterministic Markov processes}
\author[1,2,*]{Yen Ting Lin}
\author[3,4,5]{Nicolas E.~Buchler}
\affil[1]{Theoretical Division and Center for Nonlinear Studies, Los Alamos National Laboratory, New Mexico 87545, USA}	
\affil[2]{Previous affiliation: School of Physics and Astronomy, The University of Manchester, Manchester M13 9PL, UK}
\affil[3]{Department of Physics, Duke University, Durham, North Carolina 27708, USA}
\affil[4]{Department of Biology, Duke University, Durham, North Carolina 27708, USA} 
\affil[5]{Center for Genomic \& Computational Biology, Durham, North Carolina 27710, USA}
\affil[*]{Corresponding author:  yentingl@lanl.gov}
\date{\today}

\twocolumn[
  \begin{@twocolumnfalse}
	\maketitle
    \begin{abstract}
{Single-cell experiments show that gene expression is stochastic and bursty, a feature that can emerge from slow switching between promoter states with different activities. One source of long-lived promoter states is the slow binding and unbinding kinetics of transcription factors to promoters, i.e. the non-adiabatic binding regime. Here, we introduce a simple analytical framework, known as a piecewise deterministic Markov process (PDMP), that accurately describes the stochastic dynamics of gene expression in the non-adiabatic regime. We illustrate the utility of the PDMP on a non-trivial dynamical system by analyzing the properties of a titration-based oscillator in the non-adiabatic limit. We first show how to transform the underlying Chemical Master Equation into a PDMP where the slow transitions between promoter states are stochastic, but whose rates depend upon the faster deterministic dynamics of the transcription factors regulated by these promoters. We show that the PDMP accurately describes the observed periods of stochastic cycles in activator and repressor-based titration oscillators. We then generalize our PDMP analysis to more complicated versions of titration-based oscillators to explain how multiple binding sites lengthen the period and improve coherence. Last, we show how noise-induced oscillation previously observed in a titration-based oscillator arises from non-adiabatic and discrete binding events at the promoter site.}
	\end{abstract}
\end{@twocolumnfalse}
\vspace{10pt}
]

\section{Introduction}
Gene expression is fundamentally a stochastic biochemical process that arises from thermal fluctuations. An important source of stochastic noise comes from the discrete and random binding and unbinding events between the regulating transcription factors and the promoter sites of the regulated genes. Conventionally, these DNA binding and unbinding events are thought to be fast compared to the downstream processes of transcription, translation, and degradation \cite{buchler2003schemes}. This separation of timescales leads to an approximation, known as a quasi-steady state or adiabatic approximation, where the mean transcription rate simplifies to a function of the concentrations and protein-DNA dissociation constants at the promoter \cite{hornos2005self,ackers1982quantitative}. The adiabatic approximation is commonly used to reduce the number of dynamical variables (e.g. promoter states) in gene regulatory networks.  However, it is also a bold assumption because experiments \cite{kwon2001determination,kyo2004evaluation,geertz2012massively,hammar2014direct} show that the binding and unbinding events of transcription factors and chromatin at the promoter can take place at a comparable, or even slower, timescale than the downstream processes of gene expression. This observation has motivated theoretical studies into the effects of slow or non-adiabatic binding on gene regulatory networks. There is a consensus that non-adiabatic binding results in bursty production of transcripts \cite{walczak2012analytic,bressloff2017stochastic}, broadened distributions of gene expression \cite{thomas2014phenotypic,hufton2016intrinsic,jia2017simplification}, and bi- or multi-stabilities that reflect the discrete, underlying promoter states \cite{hufton2016intrinsic,jia2017simplification, al2017multi}.

\begin{figure*}[t]
\begin{center}
\includegraphics[width=0.70\textwidth]{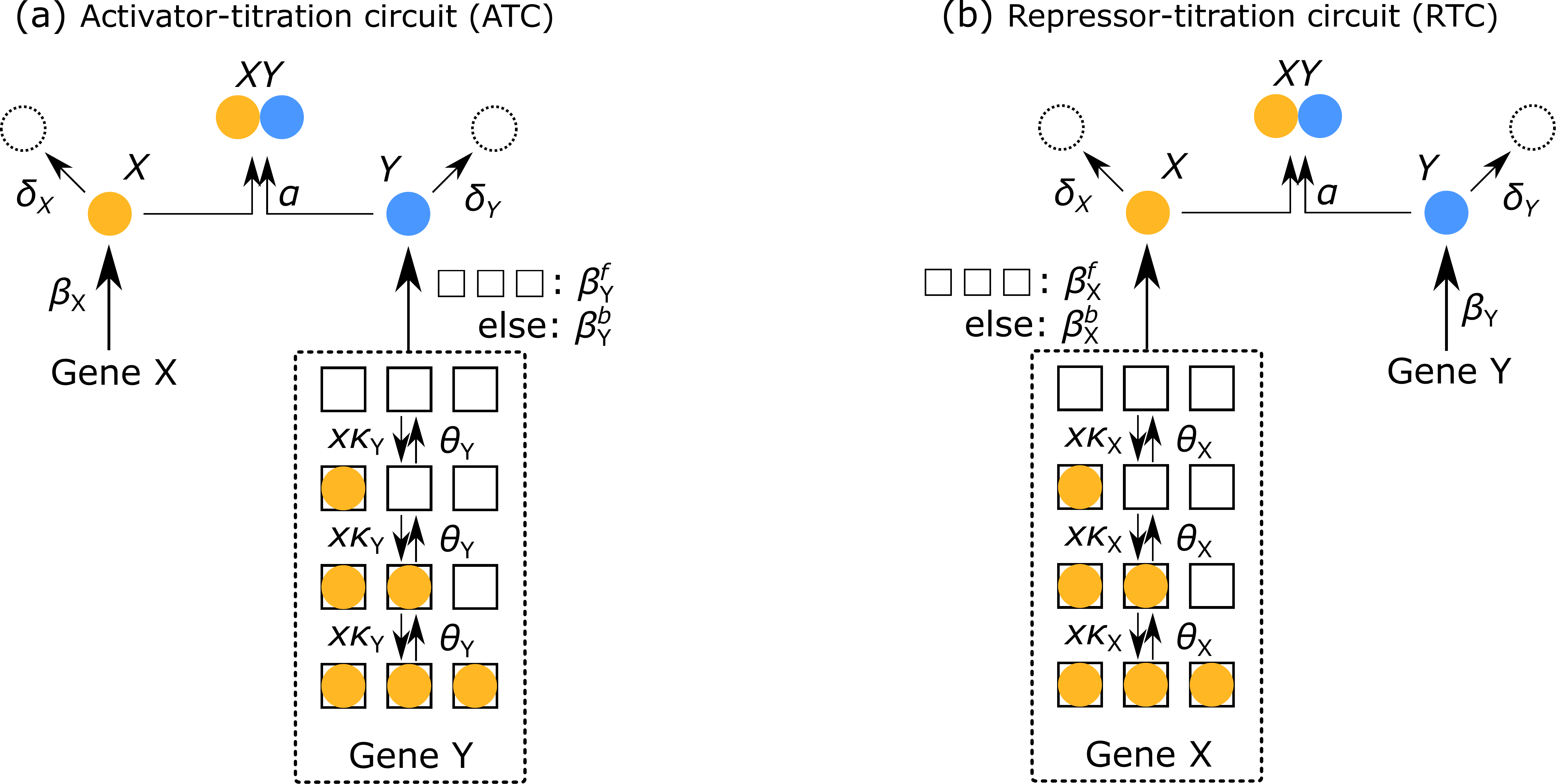}
\end{center}
\caption{Schematic diagrams of the idealised (a) activator-titration circuit (ATC) and (b) repressor-titration circuit (RTC).}
\label{fig:schematicIdealised}
\end{figure*}

Many of these studies focused on the system properties at stationarity and mostly ignored the effects of non-adiabatic binding on the non-equilibrium dynamics of gene regulatory networks. In this article, we address the following questions: What are the \emph{dynamical} consequences of non-adiabatic binding?  What kind of modeling framework accurately describes the non-stationary dynamics of gene regulatory networks in the non-adiabatic regime?  To answer these questions, we use a model of titration-based clocks to illustrate the effects of non-adiabatic binding on dynamics (e.g. oscillation) and to show how an analytical framework, known as a piecewise deterministic Markov process (PDMP), accurately describes the stochastic dynamics of the full model in the non-adiabatic regime. 

This article is organised as follows. In Section \ref{sec:idealised}, we introduce two idealised models of titration-based circuits commonly found in circadian clocks and immune signaling. We prove in Section ~\ref{sec:NoExistence} that limit cycles are impossible in the fast-binding (adiabatic) limit. In Section ~\ref{sec:ATCRTC}, we simulate the Full Chemical Master Equation to demonstrate that the titration-based circuits exhibit stochastic cycles in the slow-binding (non-adiabatic) limit.  We then transform the Chemical Master Equation into a piecewise deterministic Markov process (PDMP) where transitions between discrete promoter states are stochastic but the rates depend upon the faster deterministic dynamics of the transcription factor concentrations regulated by these promoters (Section ~\ref{sec:PDMP}). The PDMP makes no assumptions regarding the timescales of promoter switching and is valid for both slow or fast switching. It is an exact formulation of the Chemical Master Equation in the thermodynamic limit for systems where protein numbers are large. The thermodynamic limit and, hence, PDMP analysis is well-suited for stochastic gene dynamics in eukaryotic cells where cell sizes and the number of regulatory proteins can be large. We show that the PDMP framework accurately describes the observed periods and coherence of stochastic cycles in the non-adiabatic regime. We also demonstrate that the PDMP can be readily applied to more detailed and mechanistic models in Section ~\ref{sec:published}. We conclude in Section ~\ref{sec:discussion} by discussing our results and PDMP analysis in the context of previous work on non-adiabatic binding and oscillation in gene networks.

\section{Mathematical Framework}
We begin by introducing two idealised models of titration-based gene regulatory networks commonly found in biological oscillators.  These models are `idealised' in the sense that transcription and translation are lumped into a single-stage of `production', and the intermediate mRNA populations are not explicitly modelled. We further simplify the cis-regulatory architecture of each promoter to the fewest number of possible binding states. The purpose of these idealised models is to illustrate how PDMP analysis can be used to understand the origin and properties of the stochastic cycles that emerge in the non-adiabatic regime. We will relax some of these assumptions in later sections.

\subsection{Idealised models}\label{sec:idealised}

Both idealised models consist of two genes, which produce two kinds of regulatory proteins $X$ (a transcription factor, TF) and $Y$ (an inhibitor that titrates $X$ into an inactive complex); see Figure \ref{fig:schematicIdealised}. Our first model is called the activator-titration circuit (ATC) because protein $X$ is a transcriptional activator \cite{franccois2005core,karapetyan2015role}. In this model, $X$ increases the production rate of inhibitor $Y$ by binding to cis-regulatory binding sites in the promoter of gene Y with an association rate $\kappa_\text{Y}$. There are a total of $\mathcal{N}_\text{Y}$ cis-regulatory binding sites in the promoter of gene Y and we assume that binding of X is sequential, such that there are a total of $\mathcal{N}_\text{Y} + 1$ promoter states.  Bound $X$ dissociates sequentially from each binding sites with a rate $\theta_\text{Y}$. The production rate of gene Y depends non-linearly on the number of $X$ bound to the promoter because the production rate is $\beta^b_\text{Y}$ (``bound’’) when {\em any} of the binding sites are occupied; otherwise, the production rate is $\beta^f_\text{Y}$ (``free’’). We note that $\beta^b_\text{Y}>\beta^f_\text{Y}$ because $X$ is an activator. Gene X is unregulated and, thus, activator $X$ is constitutively produced at a constant rate $\beta_\text{X}$. Last, inhibitor $Y$ inhibits the activity of TF $X$ by titration, where one $Y$ molecule irreversibly binds to one $X$ molecule with a bimolecular rate of association ($\alpha$) and forms a non-functional heterodimer. The idealised ATC can be modelled by the following elementary reactions:
\al{
\varnothing \xrightarrow{\beta_\text{X}}{}& X, \text{ (Production of $X$)} \\
\varnothing \xrightarrow{\beta^f_\text{Y}}{}& Y, \text{ if } s_\text{Y}=0, \text{ (Production of $Y$)} \\
\varnothing \xrightarrow{\beta^b_\text{Y}}{}& Y, \text{ if } s_\text{Y}>0, \text{ (Production of $Y$)} \\
X \xrightarrow{\delta_X} {}&  \varnothing, \text{ (Degradation of $X$)} \\
Y \xrightarrow{\delta_Y} {}&  \varnothing, \text{ (Degradation of $Y$)} \\
X+Y \xrightarrow{\alpha} {}&  \varnothing, \text{ (Titration)}\\
s_\text{Y} \xrightarrow{x \kappa_\text{Y}}{}& s_\text{Y}+1, \text{ if } 0\le s_\text{Y} < \mathcal{N}_\text{Y}, \text{ (Binding)}\\
s_\text{Y} \xrightarrow{\theta_\text{Y}}{}& s_\text{Y}-1, \text{ if } 0 < s_\text{Y}  \le\mathcal{N}_\text{Y}. \text{ (Unbinding)}
}{eq:ATCprocess}
Here, $s_\text{Y}=0,1,\ldots \mathcal{N}_\text{Y}$ identifies the promoter state by its number of bound $X$. A schematic diagram of the ATC can be found in Fig.~\ref{fig:schematicIdealised}(a). All the elementary rate constants are defined in the sense of mass action kinetics, and $x$ denotes the concentration of TF $X$ in the thermodynamic limit. In the stochastic models that consider discrete molecules (detailed in Sec.~\ref{sec:ATCRTC}), the rates have to be properly scaled by $\Omega$, which is a parameter that quantifies the system size and is related to cell volume. The scaling relationship between the mass-action rate constants and the stochastic model rates can be found in Appendix \ref{app:scaling}.  

The second model is called a repressor-titration circuit (RTC) because $X$ is a transcriptional repressor \cite{karapetyan2015role}; see Figure \ref{fig:schematicIdealised}(b). This model differs from the ATC in two ways. First, the inhibitor $Y$ is now constitutively expressed at a constant rate $\beta_Y$. Secondly, $X$ negatively auto-regulates itself where $X$ decreases its own production rate by binding to cis-regulatory binding sites in the promoter of gene X with an association rate $\kappa_X$.  There are a total of $\mathcal{N}_\text{X}$ cis-regulatory binding sites in the promoter of gene X and we assume that binding of X is sequential, such that there are a total of $\mathcal{N}_\text{X} + 1$ promoter states. Bound $X$ dissociates sequentially from each binding sites with a rate $\theta_\text{X}$. The production rate of gene X depends non-linearly on the number of $X$ bound to the promoters, where the production rate of $X$ is $\beta^b_\text{X}$ (``bound’’) when {\em any} of the binding sites are occupied; otherwise, the production rate is $\beta^f_\text{X}$ (``free’’). We note that $\beta^f_\text{Y}>\beta^b_\text{Y}$ because $X$ is a repressor. The rest of the process and parameters are similarly defined as in the ATC. The idealised RTC can be modelled by the following elementary reactions:
\al{
\varnothing \xrightarrow{\beta_\text{Y}}{}& Y, \text{ (Production of $Y$)} \\
\varnothing \xrightarrow{\beta^f_\text{X}}{}& X, \text{ if } s_\text{X}=0, \text{ (Production of $X$)} \\
\varnothing \xrightarrow{\beta^b_\text{X}}{}& X, \text{ if } s_\text{X}>0, \text{ (Production of $X$)} \\
X \xrightarrow{\delta_X} {}&  \varnothing, \text{ (Degradation of $X$)} \\
Y \xrightarrow{\delta_Y} {}&  \varnothing, \text{ (Degradation of $Y$)} \\
X+Y \xrightarrow{\alpha} {}&  \varnothing, \text{ (Titration)}\\
s_\text{X} \xrightarrow{x \kappa_\text{X}}{}& s_\text{X}+1, \text{ if } 0\le s_\text{X} < \mathcal{N}_\text{X}, \text{ (Binding)}\\
s_\text{X} \xrightarrow{\theta_\text{X}}{}& s_\text{X}-1, \text{ if } 0 < s_\text{X}  \le\mathcal{N}_\text{X}. \text{ (Unbinding)}
}{eq:RTCprocess}
Similarly, $s_\text{X}=0,1,\ldots \mathcal{N}_\text{X}$ identifies the promoter state by its number of bound $X$. 

\subsection{No limit cycle in the adiabatic limit}\label{sec:NoExistence}
The aim of this section is to show that mass action kinetics describing the ATC and RTC in the fast-switching (adiabatic) limit do not allow deterministic limit cycles. Below, we generically use $\text{Z} = $ X or Y as the gene index and $Z=$ $X$ or $Y$ as the protein index. The discrete switching events between the bound TF at the promoter sites, $s_\text{Z} \rightarrow s_\text{Z}\pm1$, are a random birth-and-death process \cite{van1992stochastic,gardiner1985handbook} where the birth rate $\kappa_\text{Z}$ depends on the concentration ($x$) of the transcription factor $X$. In the fast-switching (adiabatic) limit, formally expressed as $\mathcal{O} \l(x \kappa_\text{Z}\r), \mathcal{O}\l(\theta_\text{Z}\r) \gg \mathcal{O}\l(\text{Any other transition rates}\r)$, the variable $x$ is a slow variable and is treated as approximately constant. In this limit, the birth and death rates are approximately constant, and the quasi-stationary distribution (QSD) of $s_\text{Z}$ is obtained using detailed balance of this one-dimensional birth-death process \cite{van1992stochastic}:
\eq{
\mathbb{P}^\text{QSD} \l\{s_\text{Z}=i\r\} =  \frac{\l(x \kappa_\text{Z} / \theta_\text{Z} \r)^i}{\sum_{m=0}^{\mathcal{N}_\text{Z}} \l(x \kappa_\text{Z} / \theta_\text{Z} \r)^{m}}.
}{eq:quasiStationary}
The effective production rate of the regulated gene, $\beta^{\rm eff}_\text{Z}$, can be derived using the quasi-stationary distribution \eqref{eq:quasiStationary}:
\eq{
\beta^{\rm eff}_\text{Z}\l(x\r) =  \beta^b_\text{Z}  + \frac{(\beta^f_\text{Z}- \beta^b_\text{Z})} {\sum_{m=0}^{\mathcal{N}_\text{Z}} \l(x \kappa_\text{Z} / \theta_\text{Z} \r)^{m}}.
}{eq:effectiveBirth}

In the thermodynamic limit, we denote the concentrations of $X$ and $Y$ by $x$ and $y$ respectively, and the resulting mass action kinetics of the ATC and RTC are described by the following deterministic differential equations:
\subeq{
\dot{x}(t) ={}& \mathcal{F}(x,y) = \beta^{\rm eff}_\text{X}\l(x\r) - \delta_X x - \alpha x y,\\
\dot{y}(t) ={}& \mathcal{G}(x,y) = \beta^{\rm eff}_\text{Y}\l(x\r) - \delta_Y y - \alpha x y.
}{eq:ODE}
For simplicity, we unified the expressions for the idealised ATC and RTC where $ \beta^{\rm eff}_\text{X}\l(x\r) :=constant~\beta_\text{X}$ in the ATC and $\beta^{\rm eff}_\text{Y}\l(x\r):=constant~\beta_\text{Y}$ in the RTC. Equations \eqref{eq:ODE} constitute a two-dimensional dynamical system. The Bendixson criterion \cite{khalil1996nonlinear} states that limit cycles do not exist when the trace of the Jacobian, $\partial_x \mathcal{F}(x,y)+\partial_y \mathcal{G}(x,y)$, does not change sign on a simply connected domain. On the biologically relevant domain $x\ge0$ and $y\ge0$,
\eq{
\partial_x \mathcal{F}+\partial_y \mathcal{G} = \frac{\dd \beta^{\rm eff}_\text{X}\l(x\r)}{\dd x}-\delta_X-\delta_Y - \alpha \l(x+y\r)<0.
}{}
The trace of the ATC is always negative because ${\dd \beta^{\rm eff}_\text{X}\l(x\r)}/{\dd x}=0$. The trace of the RTC is also always negative because ${\dd \beta^{\rm eff}_\text{X}\l(x\r)}/{\dd x}<0$. Thus, there are no deterministic limit cycles for the idealised ATC and RTC in the adiabatic limit. If we were to modify the ATC such that the activator $X$ also stimulates its own production (i.e. positive feedback), then ${\dd \beta^{\rm eff}_\text{X}\l(x\r)}/{\dd x}>0$ and it would be possible to have limit cycles in the adiabatic limit. 

\subsection{Stochastic cycles in the non-adiabatic regime} \label{sec:ATCRTC}
We first develop a full stochastic model that describes the dynamics where the population of molecules and the number of bound promoter sites are all discrete. We will then use this model to show the emergence of stochastic cycles with well-defined periods in the non-adiabatic regime. The state of the model is determined by (1) the population of $X$, $N_X$, (2) the population of $Y$, $N_Y$, (3) the bound promoter state of gene X, $s_\text{X}$, and (4) the bound promoter state of gene Y, $s_\text{Y}$. The probability of having $N_X=i$, $N_Y=j$, $s_\text{X}=k$, and $s_\text{Y}=l$ at time $t$ is given by $P_{i,j,k,l}(t)$. The discrete-state stochastic process of the idealised model is described by the Chemical Master Equation \cite{van1992stochastic,gardiner1985handbook}:
\al{
\dot{P}_{i,j,k,l} ={}& \l(P_{i-1,j,k,l}- P_{i,j,k,l}\r) \mathbf{1}_{\l\{k=0\r\}} \Omega \beta_\text{X}^f \nonumber \\
{}& +  \l( P_{i-1,j,k,l}- P_{i,j,k,l}\r) \mathbf{1}_{\l\{k>0\r\}} \Omega \beta_\text{X}^b \nonumber \\
{}& +\l(P_{i,j-1,k,l}- P_{i,j,k,l}\r) \mathbf{1}_{\l\{l=0\r\}} \Omega \beta_\text{Y}^f \nonumber \\
{}& +\l( P_{i,j-1,k,l}- P_{i,j,k,l} \r) \mathbf{1}_{\l\{l>0\r\}} \Omega \beta_\text{Y}^b \nonumber \\
{}& +\delta_X  \l[\l(i+1\r)P_{i+1,j,k,l}- i P_{i,j,k,l}\r] \nonumber\\
{}& +\delta_Y  \l[ \l(j+1\r)P_{i,j+1,k,l}- j P_{i,j,k,l} \r]\nonumber\\
{}& + \frac{\alpha}{ \Omega} \l[\l(i+1\r)\l(j+1\r)P_{i+1,j+1,k,l}-ijP_{i,j,k,l}   \r]\nonumber\\
{}& + \frac{\kappa_\text{X}}{ \Omega}  \l[\l(i+1\r) P_{i+1,j,k-1,l} - i \mathbf{1}_{\l\{k<\mathcal{N}_\text{X}\r\}} P_{i,j,k,l}\r] \nonumber \\
{}& +\frac{\kappa_\text{Y}}{ \Omega}\l[\l(i+1\r) P_{i+1,j,k,l-1}- i \mathbf{1}_{\l\{l<\mathcal{N}_\text{Y}\r\}} P_{i,j,k,l} \r]\nonumber\\
{}& +\theta_\text{X}  \l[P_{i-1,j,k+1,l} -  \mathbf{1}_{\l\{k>0\r\}} P_{i,j,k,l}\r] \nonumber \\
{}&+\theta_\text{Y}\l[P_{i+1,j,k,l+1} - \mathbf{1}_{\l\{l>0\r\}}P_{i,j,k,l}\r].
}{eq:simpleMaster}
where we have suppressed writing the $t$-dependence of $P_{i,j,k,l}$ for brevity. The boundary conditions $P_{i,j,k,l}=0$ when $i<0$, $j<0$, $k<0$, $l<0$, $k>\mathcal{N}_\text{X}$, or $l>\mathcal{N}_\text{Y}$ are imposed. We unified the model descriptions of ATC and RTC; for ATC, $\mathcal{N}_\text{X}$, $\kappa_X$, $\theta_X:=0$ and $\beta^b_\text{X}=\beta^f_\text{X}=constant~\beta_\text{X}$; similarly, for RTC, $\mathcal{N}_\text{Y}$,$\kappa_Y$, $\theta_Y:=0$ and $\beta^b_\text{Y}=\beta^f_\text{Y}=constant~\beta_\text{Y}$. $\mathbf{1}_{\l\{\text{condition}\r\}}$ is the characteristic function: it is equal to $1$ when the condition is true, otherwise $0$. The different rates and the protein population scale as a function of system size $\Omega$, as outlined in Appendix A \cite{hufton2016intrinsic,van1992stochastic}. The concentrations $x$ and $y$ in previous sections are interpreted as the normalised population density $N_X/\Omega$ and $N_Y/\Omega$, where $N_X$ and $N_Y$ are the discrete populations of regulatory proteins $X$ and $Y$. 

\begin{figure}[!t]
\begin{center}
\includegraphics[width=0.47\textwidth]{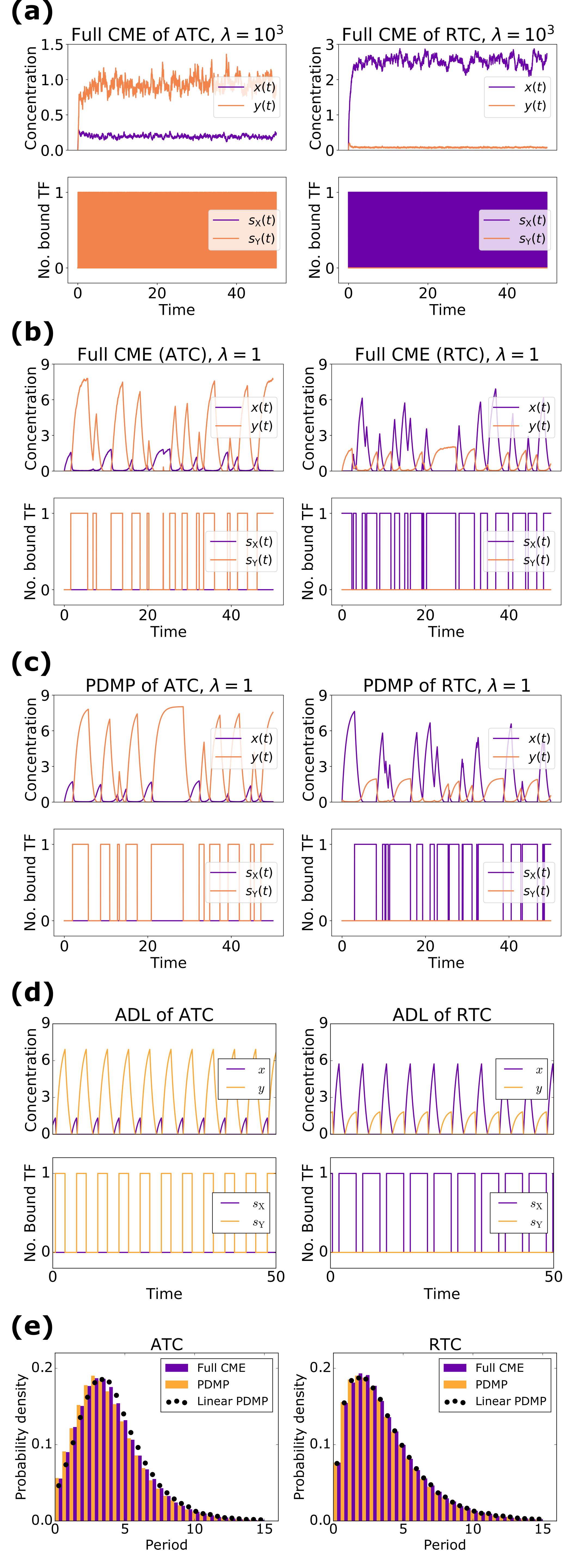}
\end{center}
\caption{Sample paths of the Full CME of the ATC and RTC in the (a) adiabatic regime ($\lambda=1000$) and (b) non-adiabatic regime ($\lambda=1$) for a single binding site ($\cal{N}$=1). (c) Sample paths of the constructed piecewise deterministic Markov process when $\lambda=1$ (Sec.~\ref{sec:PDMP}). (d) The alternative deterministic limit of the processes (Sec.~\ref{sec:dATCRTC}). (e) Quantification of the periods of the stochastic cycles.}
\label{fig:ATCRTC}
\end{figure}

We refer to the model \eqref{eq:simpleMaster} as the Full CME. Standard continuous time Markov chain simulations were constructed to generate exact sample paths of the Full CME of the ATC and RTC \cite{gillespie1976general,gillespie1977exact}. We chose two sets of parameters listed in Table \ref{table:ATCRTC}, where the ATC and RTC promoters have a single binding site ($\cal{N}_\text{Z}$=1) and the only source of nonlinearity is the titration of $X$ by $Y$. We chose this parameter set because it is simple and it illustrates the fundamental ingredients of stochastic cycling in the non-adiabatic regime.  We will consider more complicated cis-regulatory promoters in later sections.  For each parameter set, we introduce a scaling factor $\lambda$, such that the binding and unbinding rates are respectively parametrised by $\kappa_\text{Z}:=\lambda\bar{\kappa}_\text{Z}$ and $\theta_\text{Z}:=\lambda\bar{\theta}_\text{Z}$. We fixed $\bar{\kappa}_\text{Z}$ and $\bar{\theta}_\text{Z}$ and systematically change the value of $\lambda$ in order to examine the dynamics of the same model in both the adiabatic and non-adiabatic regime.

\begin{figure*}[t]
\centering
\footnotesize
\tikzstyle{free}=[rectangle,thick,minimum size=0.5cm,draw=blue!80,fill=blue!20]
\tikzstyle{bound}=[rectangle, thick, minimum size=0.5cm,draw=red!80,fill=red!20]
\tikzstyle{dumb}=[rectangle, thick, minimum size=0.5cm,draw=white!100,fill=white!100]
\begin{tikzpicture}[auto, outer sep=3pt,node distance=4.2cm,scale=0.8, every node/.style={transform shape},>=latex']
    
\node [free] (Sx00) {$\begin{array}{c} s_\text{Y}=0 \\ \dot x = \beta_\text{X}- \delta_X x - \alpha x y \\ \dot y = \beta_\text{Y}^f - \delta_Y y - \alpha x y\end{array}$};
\node [bound, right of = Sx00] (Sx30) {$\begin{array}{c} s_\text{Y}=1 \\ \dot x = \beta_\text{X} - \delta_X x - \alpha x y\\ \dot y = \beta_\text{Y}^b - \delta_Y y - \alpha x y\end{array}$};
\node [dumb, above right =  0.0cm and -0.1 cm of Sx00] (Sx80) {\large (a)};
\draw[transform canvas={yshift=0.3ex},-left to,thick] (Sx00) -- node[above]{$\kappa_\text{Y} x $}  (Sx30);
\draw[transform canvas={yshift=-0.3ex},left to-,thick] (Sx00) -- node[below] {$\theta_\text{Y} $} (Sx30); 

\node [free,below = 1.75cm of Sx00] (Sx0) {$\begin{array}{c} s_\text{Y}=0 \\ (x=0,y>0 )\\ \dot x = 0 \\ \dot y = \beta_\text{Y}^f - \beta_\text{X}- \delta_Y y\end{array}$};
\node [bound, right of = Sx0] (Sx3) {$\begin{array}{c} s_\text{Y}=1\\ ( x=0,y\ge0) \\ \dot x = 0 \\ \dot y = \beta_\text{Y}^b - \beta_\text{X}  - \delta_Y y \end{array}$};
\node [free, below of = Sx0, node distance=6cm] (Sx4) {$\begin{array}{c} s_\text{Y}=0 \\ ( x\ge0,y=0) \\ \dot x =\beta_\text{X} - \beta_\text{Y}^f  - \delta_X x  \\ \dot y = 0 \end{array}$};
\node [bound, right of = Sx4] (Sx7) {$\begin{array}{c} s_\text{Y}=1 \\ (x>0,y=0) \\ \dot x = \beta_\text{X} - \beta_\text{Y}^b  - \delta_X x  \\ \dot y = 0 \end{array}$};

\draw[<-,thick] (Sx0) -- node[above] {$\theta_\text{Y} $} (Sx3); 

\node[dumb, below right = .8 cm and .6 cm of Sx0](rotor1){};
\draw [->,line width=1.5pt,color=green](rotor1) ++(140:5mm) arc (-260:80:10mm);

\node [dumb, above right =  0.0cm and -0.05 cm of Sx0] (Sx90) {\large (c)};

\draw[->,thick] (Sx0) -- node[left] {\rotatebox{90}{\parbox[c]{5cm}{\centering Deterministic Titration}}} (Sx4); 
\draw[->,thick] (Sx0) -- node[right] {\rotatebox{90}{$\Delta t =\frac{1}{\delta_Y} \log \l(1 + \frac{\delta_Y y(t_0)}{\beta_\text{X} - \beta_\text{Y}^f}\r)$ }} (Sx4); 
\draw[->,thick] (Sx7) -- node[left] {\rotatebox{90}{\parbox[c]{5cm}{\centering Deterministic Titration}}} (Sx3); 
\draw[->,thick] (Sx7) -- node[right] {\rotatebox{90}{$\Delta t =\frac{1}{\delta_X} \log \l(1 + \frac{\delta_X x(t_0)}{\beta_\text{Y}^b-\beta_\text{X}} \r)$ }} (Sx3);

\draw[transform canvas={yshift=0.3ex},-left to,thick] (Sx4) -- node[above]{$\kappa_\text{Y} x $}  (Sx7);
\draw[transform canvas={yshift=-0.3ex},left to-,thick] (Sx4) -- node[below] {$\theta_\text{Y} $} (Sx7);

\node [free, right = 2cm of Sx30] (Sx000) {$\begin{array}{c} s_\text{X}=0 \\ \dot x = \beta_\text{X}^f- \delta_X x - \alpha x y \\ \dot y = \beta_\text{Y} - \delta_Y y - \alpha x y\end{array}$};
\node [bound, right of = Sx000] (Sx300) {$\begin{array}{c} s_\text{X}=1 \\ \dot x = \beta_\text{X}^b - \delta_X x - \alpha x y\\ \dot y = \beta_\text{Y} - \delta_Y y - \alpha x y\end{array}$};
\node [dumb, above right =  0.0cm and -0.1 cm of Sx000] (Sx800) {\large (b)};
\draw[transform canvas={yshift=0.3ex},-left to,thick] (Sx000) -- node[above]{$\kappa_\text{X} x $}  (Sx300);
\draw[transform canvas={yshift=-0.3ex},left to-,thick] (Sx000) -- node[below] {$\theta_\text{X} $} (Sx300); 

\node [free,below = 1.75cm of Sx000] (Sxp0) {$\begin{array}{c} s_\text{X}=0 \\ (x=0,y>0 )\\ \dot x = 0 \\ \dot y = \beta_\text{Y} - \beta_\text{X}^f- \delta_Y y\end{array}$};
\node [bound, right of = Sxp0] (Sxp3) {$\begin{array}{c} s_\text{X}=1 \\ ( x=0,y\ge0) \\ \dot x = 0 \\ \dot y = \beta_\text{Y} - \beta_\text{X}^b  - \delta_Y y \end{array}$};
\node [free, below of = Sxp0, node distance=6cm] (Sxp4) {$\begin{array}{c} s_\text{X}=0 \\ ( x\ge0,y=0) \\ \dot x =\beta_\text{X}^f - \beta_\text{Y}  - \delta_X x  \\ \dot y = 0 \end{array}$};
\node [bound, right of = Sxp4] (Sxp7) {$\begin{array}{c} s_\text{X}=1 \\ (x>0,y=0) \\ \dot x = \beta_\text{X}^b - \beta_\text{Y}  - \delta_X x  \\ \dot y = 0 \end{array}$};

\draw[->,thick] (Sxp0) -- node[left] {\rotatebox{90}{\parbox[c]{5cm}{\centering Deterministic Titration}}} (Sxp4); 
\draw[->,thick] (Sxp0) -- node[right] {\rotatebox{90}{$\Delta t =\frac{1}{\delta_Y} \log \l(1 + \frac{\delta_Y y(t_0)}{\beta_\text{X}^f - \beta_\text{Y}}\r)$ }} (Sxp4); 
\draw[->,thick] (Sxp7) -- node[left] {\rotatebox{90}{\parbox[c]{5cm}{\centering Deterministic Titration}}} (Sxp3); 
\draw[->,thick] (Sxp7) -- node[right] {\rotatebox{90}{$\Delta t =\frac{1}{\delta_X} \log \l(1 + \frac{\delta_X x(t_0)}{\beta_\text{Y}-\beta_\text{X}^b} \r)$ }} (Sxp3); 
\draw[transform canvas={yshift=0.3ex},-left to,thick] (Sxp4) -- node[above]{$\kappa_\text{X} x $}  (Sxp7);
\draw[transform canvas={yshift=-0.3ex},left to-,thick] (Sxp4) -- node[below] {$\theta_\text{X} $} (Sxp7); 
\draw[<-,thick] (Sxp0) -- node[above] {$\theta_\text{X} $} (Sxp3); 
\node [dumb, above right =  0.0cm and -0.05 cm of Sxp0] (Sxp90) {\large (d)};

\node[dumb, below right = 0.8 cm and .6 cm of Sxp0](rotor2){};
\draw [->,line width=1.5pt,color=green](rotor2) ++(140:5mm) arc (-260:80:10mm);
\end{tikzpicture}    

\caption{Schematic diagrams of the derived piecewise deterministic Markov process (PDMP) for (a) idealised activator-titration circuit (ATC) and (b) idealised repressor-titration circuit (RTC). Both models have a single promoter site ($\cal{N}$=1). The linearised PDMP for ATC and RTC are shown in (c) and (d) respectively, where the green arrows indicates the direction of the emergent stochastic cycles. Blue and red boxes denote promoter states with different production rates where $\beta_X > \beta_Y$ ($y \rightarrow 0$ due to titration) and $\beta_Y > \beta_X$ ($x \rightarrow 0$ due to titration), respectively.}
\label{fig:ATCRTC_schematic}
\end{figure*}
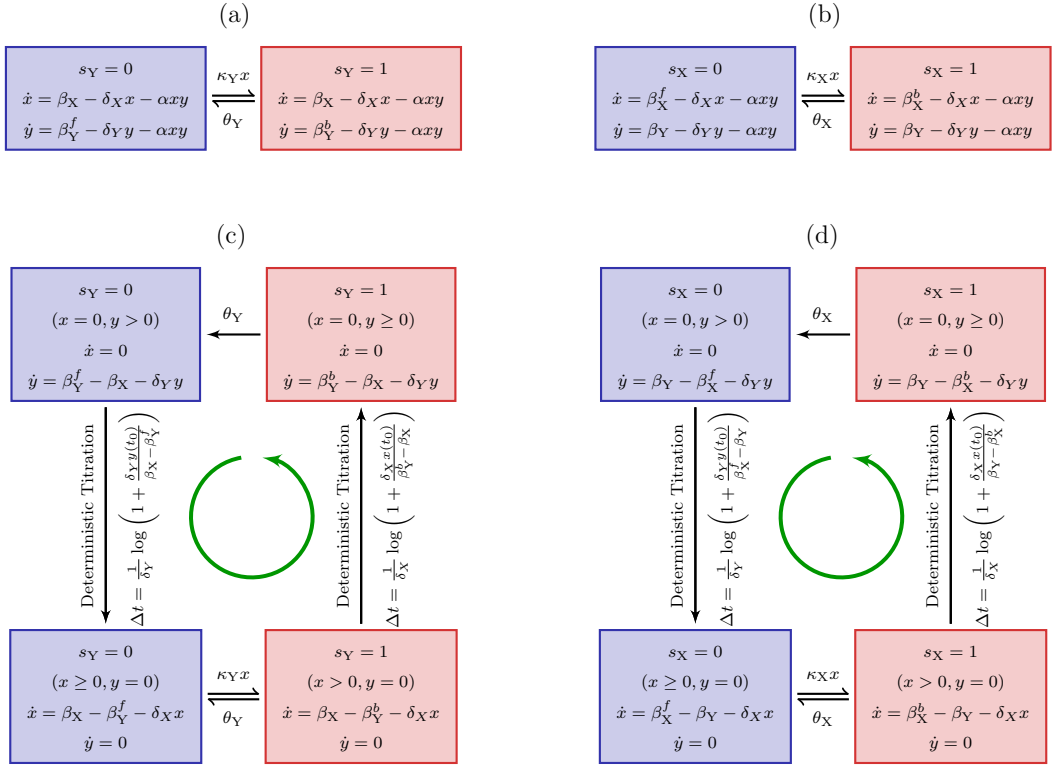

In Fig.~\ref{fig:ATCRTC}(a-b), we present sample paths of the Full CME. We do not observe limit cycles in ($x$,$y$) for the idealised ATC or RTC in the fast binding and unbinding limit (i.e., adiabatic regime, $\lambda=1000$), as predicted by our analysis in Sec.~\ref{sec:NoExistence}. When we decreased the parameter $\lambda$ to $1$, the system entered a regime where the timescale of binding and unbinding between the TF and gene is comparable to other processes. In this non-adiabatic regime, Fig.~\ref{fig:ATCRTC}(b) shows alternating high-amplitude expression of $X$ and $Y$ that appears oscillatory. We measured the `period' of each stochastic cycle using a protocol detailed in Appendix \ref{app:periods}. The measured period of stochastic cycles exhibits a unimodal distribution with a dominant frequency, as shown in Fig.~\ref{fig:ATCRTC}(e).

\subsection{Derivation of the piecewise deterministic Markov process (PDMP) approximating gene expression dynamics in the non-adiabatic regime} \label{sec:PDMP}
In this section, we develop the PDMP framework \cite{lin2016gene,lin2016bursting} of the Full CME to analyse and understand the observed stochastic cycling in the non-adiabatic regime. The idea of PDMP is to re-formulate the master equation conditioning on the discrete promoter states, $(k,l)\in \l\{0,1,\ldots \mathcal{N}_\text{X}\r\} \times \l\{0,1,\ldots \mathcal{N}_\text{Y}\r\}$. Then, for any fixed promoter states $(k,l)$, we approximate the stochastic dynamics in the TF population space using a set of ordinary differential equations, thus, leaving the discrete and Markovian stochastic switching in the $(k,l)$ space. This approximation is accurate for large system size ($\Omega$) or the thermodynamic limit \cite{kurtz1970solutions}. The PDMP framework makes no assumptions regarding relative timescales and is equally valid for adiabatic and non-adiabatic regimes in the thermodynamic limit.  To derive the PDMP, we first defined a continuum-limit probability density $p_{k,l}(x,y,t) \propto P_{i,j,k,l}(t)$ with the scaled variables $x:=i/\Omega$ and $y:=j/\Omega$. After inserting the $p_{k,l}(x,y,t)$ into the master equation \eqref{eq:simpleMaster}, performing a Kramers--Moyal expansion \cite{van1992stochastic,gardiner1985handbook} with respect to large system size $\Omega$, and collecting terms to the lowest order ($\mathcal{O} \l(\Omega^0\r)$), we arrived at the coupled partial differential equations for the probability density $p_{k,l}\equiv p_{k,l}\l(x,y,t\r)$:
\al{
\partial_t p_{k,l}={}& - \partial_x  \l[  p_{k,l} \l(\mathbf{1}_{\l\{k=0\r\}}  \beta_\text{X}^f +  \mathbf{1}_{\l\{k>0\r\}}\beta_\text{X}^b - \delta_X x \r) \r] \nonumber\\
{}&- \partial_y  \l[  p_{k,l} \l(\mathbf{1}_{\l\{l=0\r\}} \beta_\text{Y}^f +  \mathbf{1}_{\l\{l>0\r\}} \beta_\text{Y}^b - \delta_Y y  \r) \r] \nonumber \\
{}& -  \l(\partial_x + \partial_y\r)  \l(   p_{k,l} \alpha x y \r) \nonumber \\
{}&+ \kappa_\text{X} x \l( p_{k-1,l} - \mathbf{1}_{\l\{k<\mathcal{N}_\text{X}\r\}}  p_{k,l}\r) \nonumber \\
{}&+ \kappa_\text{Y} x \l( p_{k,l-1} - \mathbf{1}_{\l\{l<\mathcal{N}_\text{Y}\r\}} p_{k,l}\r) \nonumber \\
{}&+ \theta_\text{X} \l(  p_{k+1,l} -  \mathbf{1}_{\l\{k>0\r\}}  p_{k,l}\r) \nonumber \\
{}& + \theta_\text{Y} \l(  p_{k,l+1} - \mathbf{1}_{\l\{l>0\r\}}  p_{k,l}\r).
}{}
The coupled partial differential equations describe the evolution of joint probability density $p_{k,l}(x,y,t)$. 
Again, the `boundary conditions' in the $\l(k,l\r)$ space, $p_{k,l}=0$ if $k<0$, $l<0$, $k>\mathcal{N}_\text{X}$, or $l>\mathcal{N}_\text{Y}$, are imposed. Note that the evolution contains two parts: some terms contain $\partial_x$ or $\partial_y$ and describe the Liouvillian flow, whereas other terms contain $\kappa_\text{Z}$ or $\theta_\text{Z}$ and describe the Markovian switching between discrete promoter states $\l(k,l\r)$. Because the total state follows the deterministic Liouvillian flow between stochastically switching discrete state $\l(k,l\r)$, the resulting process is referred to as the piecewise deterministic Markov process (PDMP) \cite{davis1984piecewise,faggionato2009non}. The PDMP of the ATC and RTC models with a single promoter site ($\mathcal{N}=1$) are summarised in the schematic diagrams presented in Fig.~\ref{fig:ATCRTC_schematic}(a) and (b). 

Kinetic Monte Carlo simulations using the algorithm described in Appendix \ref{app:algo} were implemented to generate the sample paths of the PDMP in the non-adiabatic ($\lambda=1$) regime; see Fig.~\ref{fig:ATCRTC}(c). These PDMP sample paths capture the salient features of the dynamics of the Full CME in Fig.~\ref{fig:ATCRTC}(b). For example, the measured distribution of stochastic cycle periods using the PDMP is in perfect agreement with that of the Full CME; see Fig.~\ref{fig:ATCRTC}(e). Numerically, the advantage of the PDMP framework is that the Kinetic Monte Carlo simulations are faster than the continuous time Markov chain simulations of the Full CME because $x$,$y$ are determined by numerically-integrating ODEs, when the system size $\Omega \gg 1$.

\subsection{Linearisation of the PDMP} 
While the PDMP can be numerically simulated for any given state, the evolution of the TF concentrations is described by a set of nonlinear ordinary differential equations that do not allow for analytic solutions. The nonlinearity comes from the term $\alpha x y$, which describes the titration (second-order reaction). We developed a linearsation approximation, which utilises the fast titration limit (i.e., large $\alpha x y$ compared to any other reactions). In this limit, the molecular species ($X$ or $Y$) with the lower production rate relative to the other will be quickly sequestered and converge to almost zero concentration. As a consequence, we can separate each of the promoter states in Fig.~\ref{fig:ATCRTC_schematic}(a) and (b) into two dynamical regimes, one where $x>0$ and $y=0$ and the other where $x=0$ and $y>0$. The horizontal Markovian transitions between different $s_\text{Z}$ in each dynamical regime inherit the random switching present in the nonlinear PDMP (Fig.~\ref{fig:ATCRTC_schematic}(c) and (d)), whereas the vertical transitions between the dynamical regimes are determined solely by the faster, deterministic process of titration. In each of the compartments, we formulate a set of linear ordinary differential equations, which allow analytic solutions and facilitate quantification of the random switching times. Schematic diagrams of an ATC and RTC with a single promoter site are shown in Fig.~\ref{fig:ATCRTC_schematic}(c) and (d), and we shall refer to these models as the linearised PDMPs. We used Kinetic Monte Carlo simulations with the algorithm described in Appendix \ref{app:algo} to generate sample paths of the linear PDMP in the non-adiabatic ($\lambda=1$) regime. The measured distribution of stochastic cycle period using the linear PDMP is similar to that of the Full CME (Fig.~\ref{fig:ATCRTC}(e)), although it tended to underestimate the shorter cycles that occur in the PDMP and Full CME.

\subsection{Origin of stochastic cycles}
The PDMP schematic in Figure 3 suggests that the stochastic cycles arise from the two-state nature of the regulated promoter, which must follow cyclical Markovian dynamics ($s_\text{Z} = 0 \rightarrow 1 \rightarrow 0 \rightarrow 1 \ldots$). To demonstrate, we consider a two-state promoter with \emph{constant} transition rates ($k_+$ and $k_-$), e.g.~a promoter with a single binding site and a fixed concentration of a regulating transcription factor. This trivial two-state promoter system generates stochastic cycles with a unimodal distribution of `period' ($\tau$) given by a hypoexponential distribution: 
\begin{equation}
    \rho(\tau) = \frac{k_+ k_-}{k_+ - k_-} (e^{-k_- \tau} - e^{-k_+ \tau})
\end{equation}
with a mean period $\mu_\tau=1/k_+ + 1/k_-$ and variance $\sigma_{\tau}^2 = 1/k_+^2 + 1/k_-^2$. The mean and variance of the period are a sum of the mean and variance of the individual transitions in the two-state cycle because the waiting times are independent. The period distribution is qualitatively similar to Fig.~\ref{fig:ATCRTC}(e), which suggests that stochastic dynamics of a two-state promoter explain much of the stochastic cycling observed in the single-binding site ATC and RTC model. In the non-adiabatic regime, the faster protein dynamics faithfully track the underlying promoter state dynamics and generate large amplitude stochastic cycles in ($x$,$y$).

\begin{figure*}[h]
\begin{center}
\includegraphics[width=0.90\textwidth]{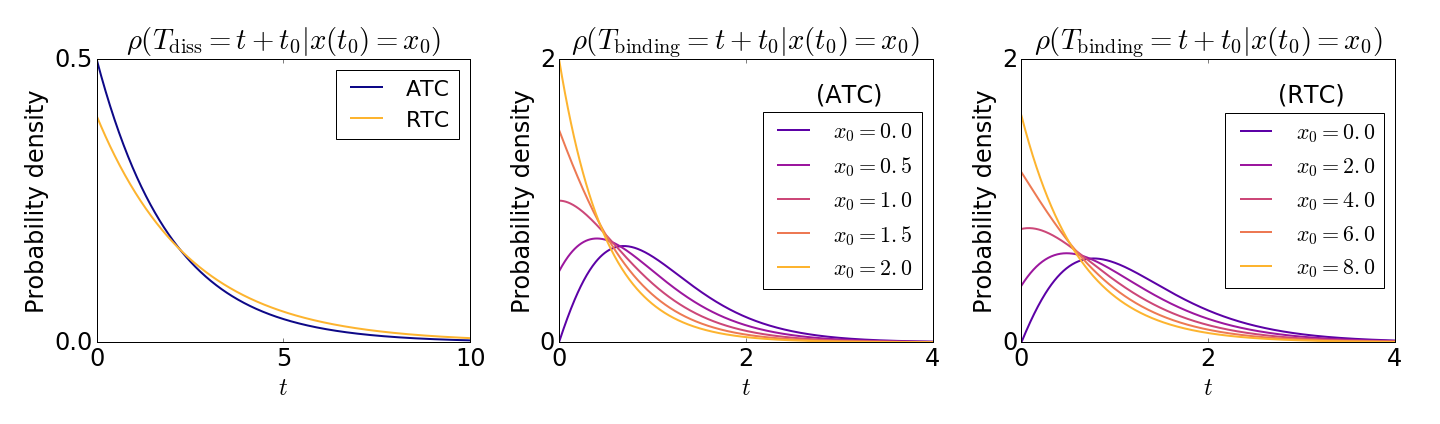}
\end{center}
\caption{
The waiting time distributions of the next dissociation and binding event in the linear PDMP. The waiting time distribution of the dissociation event is exponential. The waiting time distribution of the binding event is derived from the survival function \eqref{eq:survival}, and depends on the initial concentration of $x$.  
}
\label{fig:waitingTimes}
\end{figure*}

This raises the question of whether stochastic cycling between two promoter states can be called oscillation. This is difficult to answer because the distinction between stochastic cycling and oscillation is ill-defined. For example, by including mechanisms that reduce variance in the timing of individual transitions, one can produce cycles that are more coherent. In the extreme limit where each transition has no variance, the period of the two-state cycle has no variance and is indistinguishable from a deterministic limit cycle. 
In the following section, we will investigate potential mechanisms that reduce the variance of the stochastic transition times and make the stochastic cycles more `deterministic'.

\subsection{Increased coherence of stochastic cycles in ATC and RTC}
The linearised ATC and RTC in Figs.~\ref{fig:ATCRTC_schematic}(c) and (d) shows that the system often cycles through four discrete states that alternate between stochastic promoter switching and deterministic titration of $x$,$y$. The only state where the system has more than one `option' is $s_\text{Z}=1$ and $x>0$ (bottom right box): it can transit to either $s_\text{Z}=0$ and $x>0$ by a dissociation event of bound $X$ or to $s_\text{Z}=1$ and $x=0$ by deterministic titration of $x$. When $\theta_\text{Z}$ is sufficiently small (i.e., slow dissociation rate in the non-adiabatic regime), the system favors the later route, which induces a `full cycle' through all four discrete states in the counterclockwise order (green arrow in Fig. ~\ref{fig:ATCRTC_schematic}). As described below, this `full cycle' and the $x$-dependence of the association rate conspire to reduce variance and produce more coherent stochastic cycles.

The predominant resource of uncertainty in the `full cycle' of the ATC and RTC is the stochastic promoter switching (horizontal transitions) because the titration of $x$ (upward arrow) and $y$ (downward arrow) are deterministic and exhibit little variance. As before, the transition rate from $s_Z = 1 \rightarrow 0$ is constant and, thus, the waiting time for dissociation is a simple exponential where $\rho(t)=\theta e^{-\theta t}$; see Fig.~\ref{fig:waitingTimes}(a). Unlike the previous model, the transition from $s_Z = 0 \rightarrow 1$ is not constant and depends on $x(t)$, which can be quantified by computing the survival function \cite{cox1984analysis}. Using the linearized ATC, $x(t)$ can be exactly solved for the $s_\text{Z}=0$ state:
\eq{
x(t) = x(t_0) e^{-\delta_X \l(t-t_0\r)} + \frac{\beta_\text{X} -\beta_\text{Y}^f }{\delta_X} \l(1-e^{-\delta_X \l(t-t_0\r) }\r),
}{}
and the survival probability starting with $t=t_0$ is equal to 
\al{
\mathbb{P}{}&\l\{T_\text{binding}>t \r\} =\exp \l[ -\int_{t_0}^{t} \kappa_\text{Y} x(t') \dd t' \r ].
}{eq:survival}
The distribution of binding times is uniquely determined by this survival function, which we plot for different initial conditions $x_0$ in Fig.~\ref{fig:waitingTimes}(b). A similar calculation can be performed for the linearized RTC; see Fig.~\ref{fig:waitingTimes}(c). The variance in binding time is reduced when initial $x_0$ is close to zero because the system must wait until the population of $x$ increases to a value above which binding is likely to take place. This explains why the `full cycle' reduces variance of the total period because the system always starts at $s_\text{Z}=0$ and $x=0$ (top left box in Fig. ~\ref{fig:ATCRTC_schematic}(c-d)) due to the previous titration and dissociation of $x$. Thus, $x_0=0$ and the waiting time before $x$ binds the promoter will have reduced variance.
We remark that the binding times are not exponentially distributed and are dependent on the concentration of the activator ($x$) in general. Hence, the transition is not Markovian as some of the `memory' is stored in the TF space $(x,y)$.

\begin{figure*}[!t]
\begin{center}
\includegraphics[width=0.65\textwidth]{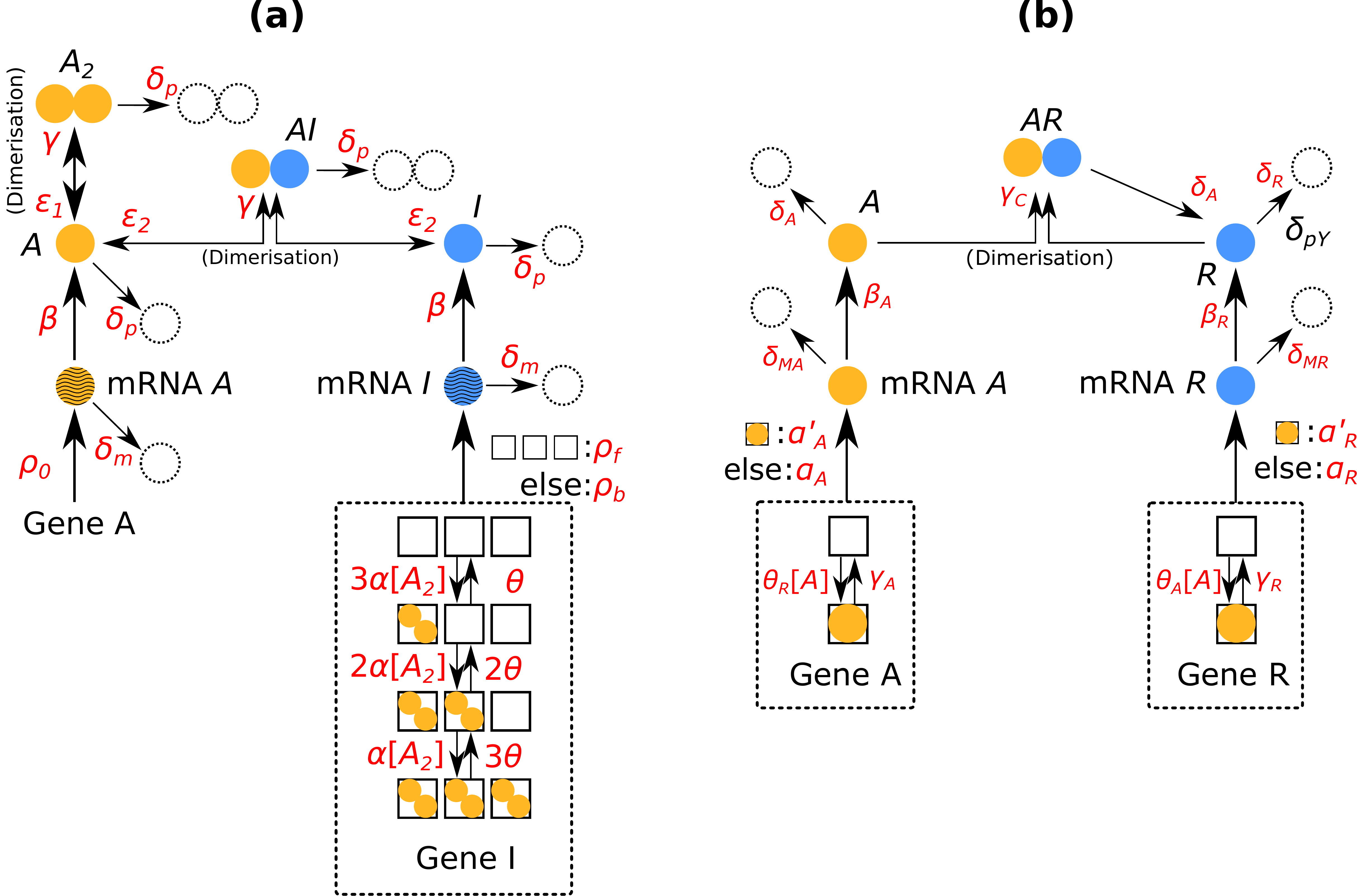}
\end{center}
\caption{
Schematic diagrams of the KB model \cite{karapetyan2015role} and the VKBL model \cite{vilar2002mechanisms} of the activator-titration circuit (ATC). The parametrisations were adopted from the original papers.
}
\label{fig:complexModels}
\end{figure*}

\subsection{Alternative deterministic limit without invoking the adiabatic approximation}\label{sec:dATCRTC}
The deterministic dynamics in Eq.~\eqref{eq:ODE} describe the mean ($x$,$y$) concentrations in the adiabatic limit where the effective protein synthesis rates are determined by the stationary distribution of promoter states. The PDMP framework explicitly models the stochastic binding and unbinding events and is valid in both the adiabatic and non-adiabatic limits. Here, we consider an alternative `deterministic limit' (ADL) of the linear PDMP, where the remaining variability due to stochastic binding and unbinding is artificially set to zero and the stochastic cycle becomes a `deterministic' limit cycle. We use the first moments of the random waiting times as a deterministic residence time of a promoter state and, thus, the dynamics in ($x$, $y$) will be deterministic. The first moments can be easily computed numerically from Eq.~\ref{eq:survival} for the linear PDMP:
\al{
\E{\Delta T_\text{binding}-t_0} = {}& - \int_{t_0}^\infty t \frac{\dd \mathbb{P}\l\{T_\text{binding}>t \r\} }{\dd t} \dd t \nonumber\\
={}&\int_{t_0}^\infty  \mathbb{P}\l\{T_\text{binding}>t \r\} \dd t.
}{}

When there is more than one possible reaction, we choose the reaction with the minimal deterministic waiting time. This is analogous to Gillespie's `first reaction' method \cite{gillespie1976general}. For the parameter set of the idealised ATC and RTC in Figure 2, the average time to titrate all the $X$ is shorter than the average dissociation time and the system cycles through all four states. The time series of the ATC and RTC in the alternative deterministic limit is shown in Fig.~\ref{fig:ATCRTC}(d).

\section{Analyses of more detailed mechanistic models } \label{sec:published}
The PDMP framework will now be applied to more sophisticated models of the ATC. A previous model of the ATC \cite{karapetyan2015role}, which we call the KB model, showed that multiple binding sites lengthened the period and improved coherence of stochastic cycling. Using the PDMP, we will show that multiple binding sites per se are insufficient to improve coherence. Rather, slow mRNA dynamics and multiple binding sites conspire to push the dynamics across all the promoter states and improve the coherence of stochastic cycling. The second model of the ATC \cite{vilar2002mechanisms}, which we call the VKBL model, has an additional positive feedback loop where the activator activates itself in addition to activating the inhibitor. The authors previously showed that the VKBL model exhibits excitation-relaxation or noise-induced oscillation beyond the Hopf bifurcation. We will use PDMP analysis to reveal that the fluctuations in the random unbinding events of the bound TF on the promoter sites are essential for inducing transcriptional noise, which in turn drive the excitable system away from its stable fixed point and induce large excursions in a semi-periodic manner. 

\subsection{Multiple binding sites do not improve coherence of stochastic cycling in idealised ATC and RTC}\label{sec:multiATC}

One explanation for the improved coherence of stochastic cycling in the KB model is that the coefficient of variance (CV) of the total period is reduced by increasing the number of steps in the `full cycle'. For example, if there are $N$ independent, stochastic steps in the full cycle and if the means and variances at each step are of equal magnitude, then the mean and variance of the total period scales with $N$ but the CV decreases as $\sqrt{N}$. To test this idea, we simulated the idealised ATC and RTC with multiple promoter sites ($\mathcal{N}_\text{Z}=3$) using the Full CME with the parameters listed in Table \ref{table:ATCRTC}. As before, we only see stochastic cycling in the non-adiabatic limit; see Fig.~\ref{fig:ATCRTC_N=3}(a-b). Strikingly, the distribution of periods was similar to that of simulations for single binding-sites; compare Fig.~\ref{fig:ATCRTC}(e) to Fig.~\ref{fig:ATCRTC_N=3}(e). To understand why multiple binding sites did not increase the period or improve the coherence of stochastic cycles, we first transformed the Full CME into a PDMP (Appendix D). We confirmed that simulations of the PMDP accurately reproduced the results of the full CME; see Fig.~\ref{fig:ATCRTC_N=3}(c). We then reduced the PDMP into a linearised PDMP framework, which explains why multiple binding sites in the idealised ATC and RTC do not significantly alter the length of the period or improve coherence. The linear PDMP shows that the system becomes trapped in a `mini-cycle' between the $s_Z=0$ and $s_Z=1$ promoter states at the blue and red boundaries (Fig.~\ref{fig:lATCRTC_schematic}). The production rate changes instantaneously upon promoter state switching across the boundary, and deterministic titration of $x$ will immediately start pushing the system upwards (red box). The timescale of titration is typically faster than that of the next stochastic binding and, thus, produces a stochastic mini-cycle around the boundary. This mini-cycle dynamic is also reflected in the alternative deterministic limit of the ATC and RTC, as shown in Fig.~\ref{fig:ATCRTC_N=3}(d).

\begin{figure*}[t]
\centering
\footnotesize
\tikzstyle{free}=[rectangle,thick,minimum size=0.5cm,draw=blue!80,fill=blue!20]
\tikzstyle{bound}=[rectangle, thick, minimum size=0.5cm,draw=red!80,fill=red!20]
\tikzstyle{dumb}=[rectangle, thick, minimum size=0.5cm,draw=white!100,fill=white!100]
\begin{tikzpicture}[auto, outer sep=3pt,node distance=4.2cm,scale=0.78, every node/.style={transform shape},>=latex']

\node [free] (Sx0) {$\begin{array}{c} s_\text{Y}=0 \\ (x=0,y>0 )\\ \dot x = 0 \\ \dot x = \beta_\text{Y}^f - \beta_\text{X}- \delta_Y y\end{array}$};
\node [bound, right of = Sx0] (Sx1) {$\begin{array}{c} s_\text{Y}=1 \\ ( x=0,y>0 ) \\ \dot x = 0 \\ \dot y = \beta_\text{Y}^b - \beta_\text{X}  - \delta_Y y \end{array}$};
\node [bound, right of = Sx1] (Sx2) {$\begin{array}{c} s_\text{Y}=2 \\ (  x=0,y>0 )\\ \dot x = 0 \\ \dot y = \beta_\text{Y}^b - \beta_\text{X}  - \delta_Y y \end{array}$};
\node [bound, right of = Sx2] (Sx3) {$\begin{array}{c} s_\text{Y}=3 \\ ( x=0,y\ge0) \\ \dot x = 0 \\ \dot y = \beta_\text{Y}^b - \beta_\text{X}  - \delta_Y y \end{array}$};
\node [free, below of = Sx0, node distance=6cm] (Sx4) {$\begin{array}{c} s_\text{Y}=0 \\ ( x\ge0,y=0) \\ \dot x =\beta_\text{X} - \beta_\text{Y}^f  - \delta_X x  \\ \dot y = 0 \end{array}$};
\node [bound, right of = Sx4] (Sx5) {$\begin{array}{c} s_\text{Y}=1 \\ ( x\ge0,y=0) \\ \dot x = \beta_\text{X} - \beta_\text{Y}^b  - \delta_X x  \\ \dot y = 0 \end{array}$};
\node [bound, right of = Sx5] (Sx6) {$\begin{array}{c} s_\text{Y}=2 \\ ( x\ge0,y=0 )\\ \dot x = \beta_\text{X} - \beta_\text{Y}^b  - \delta_X x  \\ \dot y = 0 \end{array}$};
\node [bound, right of = Sx6] (Sx7) {$\begin{array}{c} s_\text{Y}=3 \\ (x>0,y=0) \\ \dot x = \beta_\text{X} - \beta_\text{Y}^b  - \delta_X x  \\ \dot y = 0 \end{array}$};

\draw[<-,thick] (Sx0) -- node[above] {$\theta_\text{Y} $} (Sx1); 
\draw[<-,thick] (Sx1) -- node[above] {$\theta_\text{Y} $} (Sx2); 
\draw[<-,thick] (Sx2) -- node[above] {$\theta_\text{Y} $} (Sx3); 

\draw[->,thick] (Sx0) -- node[left] {\rotatebox{90}{\parbox[c]{5cm}{\centering Deterministic Titration}}} (Sx4); 
\draw[->,thick] (Sx0) -- node[right] {\rotatebox{90}{$\Delta t =\frac{1}{\delta_Y} \log \l(1 + \frac{\delta_Y y(t_0)}{\beta_\text{X} - \beta_\text{Y}^f}\r)$ }} (Sx4); 
\draw[<-,thick] (Sx1) -- node[left] {\rotatebox{90}{\parbox[c]{5cm}{\centering Deterministic Titration}}} (Sx5); 
\draw[<-,thick] (Sx1) -- node[right] {\rotatebox{90}{$\Delta t =\frac{1}{\delta_X} \log \l(1 + \frac{\delta_X x(t_0)}{\beta_\text{Y}^b-\beta_\text{X}} \r)$ }} (Sx5); 
\draw[<-,thick] (Sx2) -- node[left] {\rotatebox{90}{\parbox[c]{5cm}{\centering Deterministic Titration}}} (Sx6); 
\draw[<-,thick] (Sx2) -- node[right] {\rotatebox{90}{$\Delta t =\frac{1}{\delta_X} \log \l(1 + \frac{\delta_X x(t_0)}{\beta_\text{Y}^b-\beta_\text{X}} \r)$ }} (Sx6); 
\draw[->,thick] (Sx7) -- node[left] {\rotatebox{90}{\parbox[c]{5cm}{\centering Deterministic Titration}}} (Sx3); 
\draw[->,thick] (Sx7) -- node[right] {\rotatebox{90}{$\Delta t =\frac{1}{\delta_X} \log \l(1 + \frac{\delta_X x(t_0)}{\beta_\text{Y}^b-\beta_\text{X}} \r)$ }} (Sx3); 

\node[dumb, below right = 0.8 cm and .6 cm of Sx0](rotor2){};
\draw [->,line width=1.5pt,color=green](rotor2) ++(140:5mm) arc (-260:80:10mm);

\node[dumb, below right = 0.8 cm and .6 cm of Sx1](rotor3){};
\draw [<-,loosely dotted, line width=1.5pt,color=green](rotor3) ++(140:5mm) arc (90:-90:10mm);

\draw[transform canvas={yshift=0.3ex},-left to,thick] (Sx4) -- node[above]{$ \kappa_\text{Y} x $}  (Sx5);
\draw[transform canvas={yshift=-0.3ex},left to-,thick] (Sx4) -- node[below] {$\theta_\text{Y} $} (Sx5); 
\draw[transform canvas={yshift=0.3ex},-left to,thick] (Sx5) -- node[above]{$\kappa_\text{Y} x $}  (Sx6);
\draw[transform canvas={yshift=-0.3ex},left to-,thick] (Sx5) -- node[below] {$\theta_\text{Y} $} (Sx6); 
\draw[transform canvas={yshift=0.3ex},-left to,thick] (Sx6) -- node[above]{$\kappa_\text{Y} x $}  (Sx7);
\draw[transform canvas={yshift=-0.3ex},left to-,thick] (Sx6) -- node[below] {$\theta_\text{Y} $} (Sx7); 
\end{tikzpicture}    

\caption{Schematic diagram of the linearised PDMP describing the idealised ATC with multiple binding sites ($\mathcal{N}_\text{Y}=3$) in the non-adiabatic regime.  The green arrows indicates the emergent cycles which are predominantly observed in the simulations. The path of the dotted arrow is also observed, but less frequently (Fig.~\ref{fig:ATCRTC_N=3}(b)).}
\label{fig:lATCRTC_schematic}
\end{figure*}
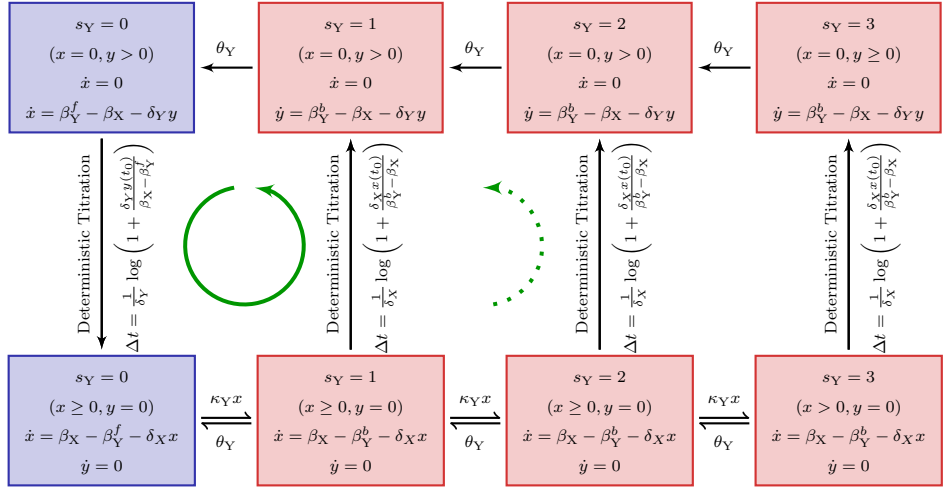
\begin{figure*}[h]
\footnotesize
\centering
\tikzstyle{free}=[rectangle,thick,minimum size=0.5cm,draw=blue!80,fill=blue!20]
\tikzstyle{bound}=[rectangle, thick, minimum size=0.5cm,draw=red!80,fill=red!20]
\tikzstyle{dumb}=[rectangle, thick, minimum size=0.5cm,draw=white!100,fill=white!100]
\begin{tikzpicture}[auto, outer sep=3pt, node distance=4.2 cm,scale=0.78, every node/.style={transform shape},>=latex']
\node [free] (Sx0) {$\begin{array}{c} G=0 \\ (a=0,z>0 )\\ \dot{r}_1 =\rho_0 -\delta_m r_1 \\ \dot{r}_2 =\rho_f -\delta_m r_2 \\ \dot{z} =\beta (r_2 - r_1)-\delta_p z  \end{array}$};
\node [bound, right of = Sx0] (Sx1) {$\begin{array}{c} G=1 \\ ( a=0,z>0 ) \\\dot{r}_1 =\rho_0 -\delta_m r_1 \\ \dot{r}_2 =\rho_b -\delta_m r_2 \\ \dot{z} =\beta (r_2 - r_1)-\delta_p z  \end{array}$};
\node [bound, right of = Sx1] (Sx2) {$\begin{array}{c} G=2 \\ (  a=0,z>0 )\\\dot{r}_1 =\rho_0 -\delta_m r_1 \\ \dot{r}_2 =\rho_b -\delta_m r_2 \\  \dot{z} = \beta(r_2 - r_1)-\delta_p z \end{array}$};
\node [bound, right of = Sx2] (Sx3) {$\begin{array}{c} G=G_{\max}=3   \\ ( a=0,z\ge0) \\ \dot{r}_1 =\rho_0 -\delta_m r_1 \\ \dot{r}_2 =\rho_b -\delta_m r_2 \\  \dot{z} =\beta(r_2 - r_1)-\delta_p z \end{array}$};
\node [free, below of = Sx0, node distance=7cm] (Sx4) {$\begin{array}{c} G=0 \\ ( a\ge0,z=0) \\  \dot{r}_1 =\rho_0 -\delta_m r_1 \\ \dot{r}_2 =\rho_f -\delta_m r_2 \\ \dot{a} = \beta (r_1 - r_2)-\delta_p a \\  \end{array}$};
\node [bound, right of = Sx4] (Sx5) {$\begin{array}{c} G=1 \\ ( a\ge0,z=0) \\ \dot{r}_1 =\rho_0 -\delta_m r_1 \\ \dot{r}_2 =\rho_b -\delta_m r_2 \\ \dot{a} = \beta (r_1 - r_2)-\delta_p a \\ \end{array}$};
\node [bound, right of = Sx5] (Sx6) {$\begin{array}{c} G=2 \\ ( a\ge0,z=0 )\\ \dot{r}_1 =\rho_0 -\delta_m r_1 \\ \dot{r}_2 =\rho_b -\delta_m r_2 \\ \dot{a} =  \beta(r_1 - r_2)-\delta_p a \\  \end{array}$};
\node [bound, right of = Sx6] (Sx7) {$\begin{array}{c} G=G_{\max}=3  \\ (a>0,z=0) \\ \dot{r}_1 =\rho_0 -\delta_m r_1 \\ \dot{r}_2 =\rho_b -\delta_m r_2 \\ \dot{a} = \beta (r_1 - r_2)-\delta_p a \\ \end{array}$};

\node[dumb, below right = 3.5 cm and .3 cm of Sx2](rotor2){};
\draw [->,line width=1.5pt,color=green](rotor2) ++(140:5mm) arc (-90:90:14.5mm);

\node[dumb, below right = 3.5 cm and .3 cm of Sx0](rotor1){};
\draw [<-,line width=1.5pt,color=green](rotor1) ++(140:5mm) arc (270:90:14.5mm);

\node[dumb, below right = 3.2 cm and -2.0 cm of Sx1](rotor3){};
\draw [->,line width=1.5pt,color=green](rotor3) -- ++(0:3.3 cm);

\node[dumb, above right = 3.1 cm and -2.0 cm of Sx5](rotor4){};
\draw [<-,line width=1.5pt,color=green](rotor4) -- ++(0:3.3 cm);

%

\draw[<-,thick] (Sx0) -- node[above] {$\theta $} (Sx1); 
\draw[<-,thick] (Sx1) -- node[above] {$2\theta $} (Sx2); 
\draw[<-,thick] (Sx2) -- node[above] {$3\theta $} (Sx3); 

\draw[->,thick] (Sx0) -- node[left] {\rotatebox{90}{\parbox[c]{5cm}{\centering Deterministic Titration}}} (Sx4); 
\draw[<-,thick] (Sx1) -- node[left] {\rotatebox{90}{\parbox[c]{5cm}{\centering Deterministic Titration}}} (Sx5); 
\draw[<-,thick] (Sx2) -- node[left] {\rotatebox{90}{\parbox[c]{5cm}{\centering Deterministic Titration}}} (Sx6); 
\draw[->,thick] (Sx7) -- node[left] {\rotatebox{90}{\parbox[c]{5cm}{\centering Deterministic Titration}}} (Sx3); 

\draw[transform canvas={yshift=0.3ex},-left to,thick] (Sx4) -- node[above]{$ 3 \alpha y $}  (Sx5);
\draw[transform canvas={yshift=-0.3ex},left to-,thick] (Sx4) -- node[below] {$\theta $} (Sx5); 
\draw[transform canvas={yshift=0.3ex},-left to,thick] (Sx5) -- node[above]{$ 2  \alpha y $}  (Sx6);
\draw[transform canvas={yshift=-0.3ex},left to-,thick] (Sx5) -- node[below] {$2 \theta $} (Sx6); 
\draw[transform canvas={yshift=0.3ex},-left to,thick] (Sx6) -- node[above]{$  \alpha y $}  (Sx7);
\draw[transform canvas={yshift=-0.3ex},left to-,thick] (Sx6) -- node[below] {$3 \theta $} (Sx7); 
\end{tikzpicture}
\caption{The linearised PDMP for the KB model \cite{karapetyan2015role}. To make the expressions compact, we define $r_1:=\l[r_A\r]$, $r_2:=\l[r_I\r]$, $z:=\l[I\r]$, and $a:=\l[\mathcal{A}\r]\equiv \l[A\r]+2\l[A_2\r]$. Green arrows indicate the direction of the emerged cycle which are frequently observed.} 
\label{fig:KBATCPDMP}
\end{figure*}
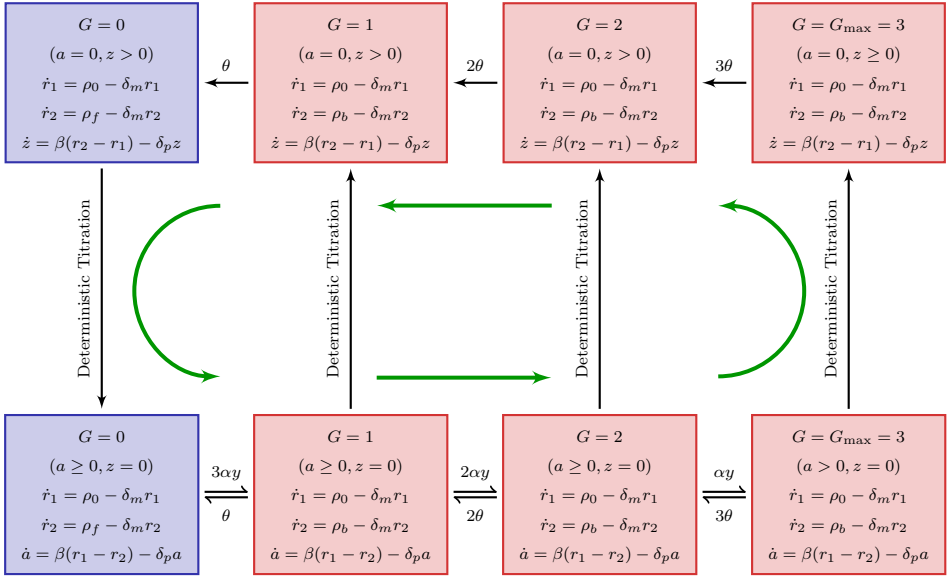

\begin{figure*}[!t]
\centering
\footnotesize
\tikzstyle{free}=[rectangle,thick,minimum size=0.5cm,draw=blue!80,fill=blue!20]
\tikzstyle{bound}=[rectangle, thick, minimum size=0.5cm,draw=red!80,fill=red!20]
\tikzstyle{dumb}=[rectangle, thick, minimum size=0cm,draw=white!100,fill=white!100]
\begin{tikzpicture}[auto, outer sep=3pt, node distance=4.2 cm,scale=0.78, every node/.style={transform shape},>=latex']
\node [free] (Sx0) {$\begin{array}{c} G=0 \\ (a=0,z>0 )\\ \dot{r}_1 =\rho_0 -\delta_m r_1 \\ \dot{r}_2 =\rho_f -\delta_m r_2 \\ \dot{z} =\beta (r_2 - r_1)-\delta_p z  \end{array}$};
\node [bound, right of = Sx0] (Sx1) {$\begin{array}{c} G=1 \\ ( a=0,z>0 ) \\\dot{r}_1 =\rho_0 -\delta_m r_1 \\ \dot{r}_2 =\rho_b -\delta_m r_2 \\ \dot{z} =\beta (r_2 - r_1)-\delta_p z  \end{array}$};
\node [bound, right of = Sx1] (Sx2) {$\begin{array}{c} G=2 \\ (  a=0,z>0 )\\\dot{r}_1 =\rho_0 -\delta_m r_1 \\ \dot{r}_2 =\rho_b -\delta_m r_2 \\ \dot{z} =\beta(r_2 - r_1)-\delta_p z \end{array}$};
\node [bound, right of = Sx2] (Sx3) {$\begin{array}{c} G=G_{\max}=3   \\ ( a=0,z\ge0) \\ \dot{r}_1 =\rho_0 -\delta_m r_1 \\ \dot{r}_2 =\rho_b -\delta_m r_2 \\ \dot{z} =\beta (r_2 - r_1)-\delta_p z \end{array}$};
\node [free, below of = Sx0, node distance=6cm] (Sx4) {$\begin{array}{c} G=0 \\ ( a\ge0,z=0) \\  \dot{r}_1 =\rho_0 -\delta_m r_1 \\ \dot{r}_2 =\rho_f -\delta_m r_2 \\ \dot{a} =\beta(r_1 - r_2)-\delta_p a \\  \end{array}$};
\node [bound, below of = Sx3, node distance =6cm] (Sx7) {$\begin{array}{c} G=G_{\max}=3  \\ (a>0,z=0) \\ \dot{r}_1 =\rho_0 -\delta_m r_1 \\ \dot{r}_2 =\rho_b -\delta_m r_2 \\ \dot{a} =  \beta(r_1 - r_2)-\delta_p a \\\end{array}$};

\draw[<-,thick] (Sx0) -- node[above] {$\theta $} (Sx1); 
\draw[<-,thick] (Sx1) -- node[above] {$2\theta $} (Sx2); 
\draw[<-,thick] (Sx2) -- node[above] {$3\theta $} (Sx3); 

\draw[->,thick] (Sx0) -- node[left] {\rotatebox{90}{\parbox[c]{5cm}{\centering Deterministic}}} (Sx4); 
\draw[->,thick] (Sx0) -- node[right] {\rotatebox{90}{\parbox[c]{5cm}{\centering Titration}}} (Sx4); 

\draw[->,thick] (Sx7) -- node[left] {\rotatebox{90}{\parbox[c]{5cm}{\centering Deterministic}}} (Sx3); 
\draw[->,thick] (Sx7) -- node[right] {\rotatebox{90}{\parbox[c]{5cm}{\centering Titration}}} (Sx3);

\node[dumb, below right = 2.9 cm and .3 cm of Sx2](rotor2){};
\draw [->,line width=1.5pt,color=green](rotor2) ++(140:5mm) arc (-90:90:10mm);

\node[dumb, below right = 2.9 cm and .3 cm of Sx0](rotor1){};
\draw [<-, line width=1.5pt,color=green](rotor1) ++(140:5mm) arc (270:90:10mm);

\node[dumb, below right = 2.58 cm and -4.0 cm of Sx1](rotor3){};
\draw [-,line width=1.5pt,color=green](rotor3) -- ++(0:8.2 cm);

\node[dumb, below right = 0.58 cm and -0.29 cm of Sx2](rotor4){};
\draw [-,line width=1.5pt,color=green](rotor4) -- ++(180:8.2 cm);
%

\draw[transform canvas={yshift=0.3ex},->,thick] (Sx4) -- node[above]{Deterministic activation}  (Sx7);
\end{tikzpicture}
\caption{The reduced PDMP approximating the KB model\cite{karapetyan2015role}. The waiting time of the deterministic activation is computed as the first moment of the cumulative distribution Eq.~\eqref{eq:survivalYay}. 
}
\label{fig:KBreduced}
\end{figure*}
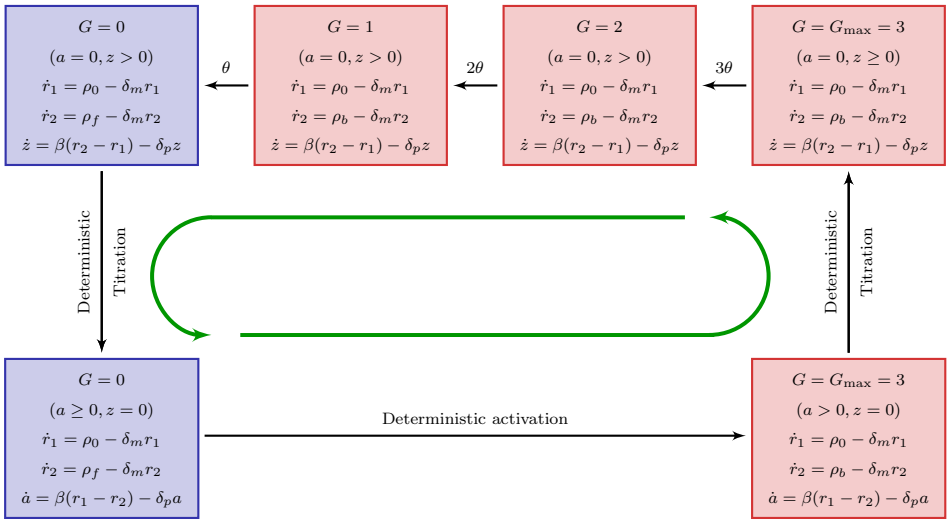

\subsection{Origins of improved coherence in the KB model}\label{sec:KB}
The KB model has several additional features compared to the idealised ATC, which could explain the observed increase in the period and coherence of stochastic cycles. First, the dynamics of mRNA transcription, degradation and protein translation are explicitly modelled. Second, the activators form homodimers before they can bind to the promoter sites and regulate gene expression. Third, the homodimers bind to the promoter sites independently and no longer need to bind sequentially (i.e. distributive binding). Last, the activator and inhibitor heterodimer is no longer irreversible and can dissociate to form monomers.

\begin{figure*}[t]
\begin{center}
\includegraphics[width=0.85\textwidth]{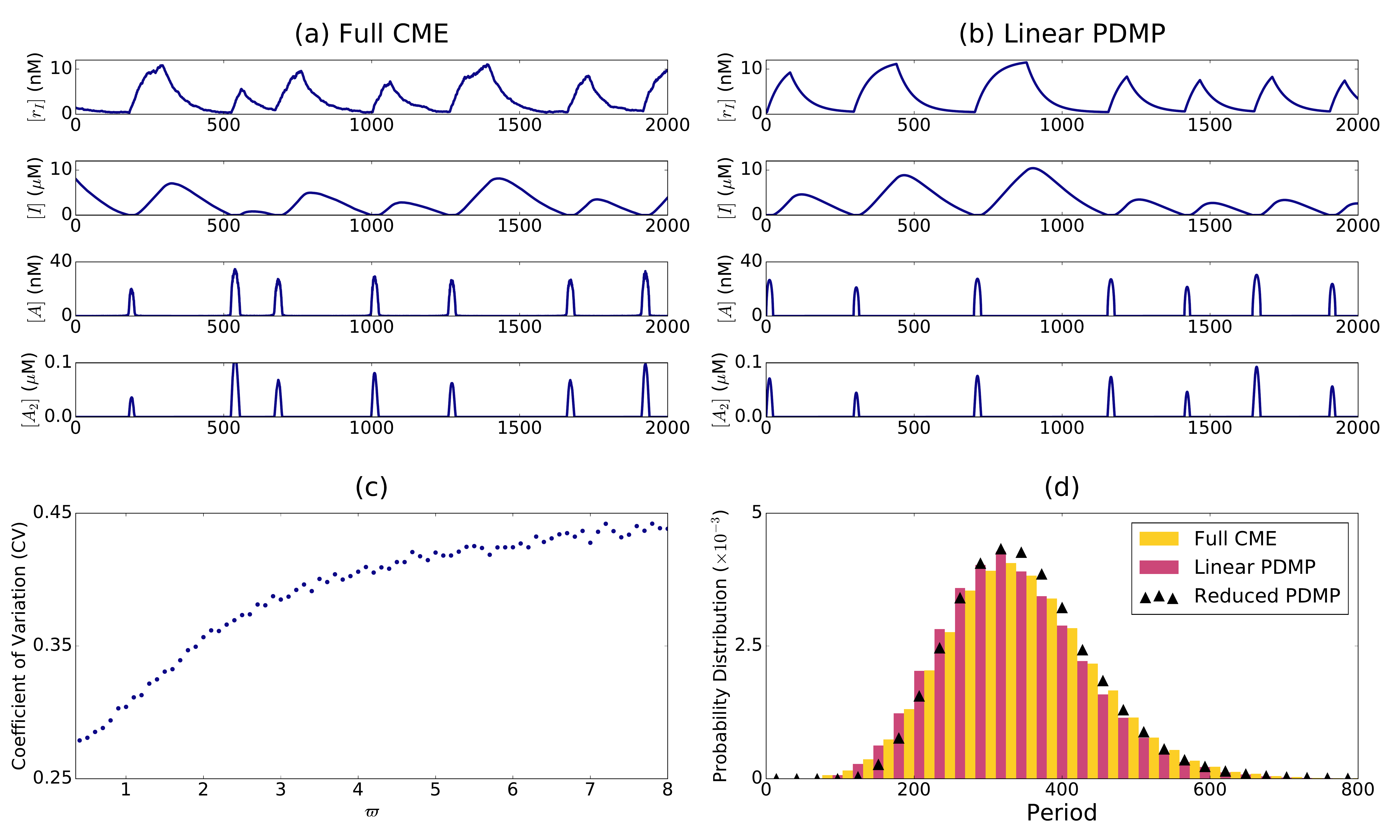}
\end{center}
\caption{
Numerically measured data of the KB model \cite{karapetyan2015role}. (a) A sample path of the Full CME, (b) a sample path of the linear PDMP (Fig.~\ref{fig:KBATCPDMP}). The stochastic cycles of the model were measured using the protocol provided in Appendix \ref{app:periods}. The full CME exhibited a minor fraction of short period cycles in the genetic states. To quantify the predominant longer periods $(100 \lesssim \text{period})$, we discarded any periods less than $50$ to generate panels (c) and (d). Panel (c) presents the coefficient of variation (CV) of the stochastic periods measured in the Full CME as a function of the scaling factor $\varpi$. Each point was computed from $10^4$ stochastic cycles. The larger the $\varpi$, the more short-lived the mRNA. The long-lived mRNA introduces a delay in protein production and improves the CV. Panel (d) presents the probability distribution of the stochastic periods when the scaling factor $\varpi=1$, as measured from $10^5$ stochastic cycles of the Full CME, PDMP, and reduced PDMP (Fig.~\ref{fig:KBreduced}).
}
\label{fig:KB}
\end{figure*}

Below, we describe PDMP analysis of the KB model \cite{karapetyan2015role} for the ATC shown in Fig~\ref{fig:complexModels}(a). Beginning with the master equation governing the KB model, we perform the system-size expansion presented in Sec.~\ref{sec:PDMP} and arrived at the following PDMP:
\al{
\frac{\dd}{\dd t} \l[{r}_A \r] {}&= - \delta_m \l[r_A\r] + \frac{\rho_0}{V}  , \nonumber\\  
\frac{\dd}{\dd t} \l[{r}_I\r] {}&=- \delta_m \l[r_I\r] + \frac{\mathbf{1}_{\l\{G=0\r\}} \rho_f + \mathbf{1}_{\l\{G>0\r\}}  \rho_b }{V}  ,\nonumber\\ 
\frac{\dd}{\dd t} \l[{A}  \r]{}& = - \delta_p \l[{A}  \r]   + \beta \l[{r}_A \r]  - 2 \gamma  \l[{A}  \r] ^2 + 2 \epsilon_1  \l[{A_2}  \r] \nonumber \\
 {}& \hphantom{={}}  - \gamma  \l[A \r]   \l[I\r] + \epsilon_2  \l[AI\r] ,\nonumber \\ 
\frac{\dd}{\dd t} \l[{A}_2  \r] {}&= -\delta_p \l[{A}_2  \r] + \gamma \l[A \r]^2 - \epsilon_1  \l[{A}_2  \r], \label{eq:KBPDMP}\\ 
\frac{\dd}{\dd t} \l[I  \r] {}&= -\delta_p  \l[I  \r]  + \beta \l[{r}_I\r]  -\gamma  \l[A \r]   \l[I\r]  + \epsilon_2  \l[AI\r], \nonumber\\
\frac{\dd}{\dd t}  \l[AI\r] {}&= -\delta_p  \l[AI\r] + \gamma  \l[A \r]   \l[I\r]   - \epsilon_2 \l[AI\r] , \nonumber\\ 
G {}&\xrightarrow{ \alpha (G_{\max}-G)\l[A_2\r]  }{G+1},\nonumber \\
G {}&\xrightarrow{ \theta G  }{G-1}\nonumber. 
}{}
We used the same variables and parameter set in Fig.~6 of the original paper \cite{karapetyan2015role}. The state variables, $\l[{r}_A \r]$, $\l[{r}_I \r]$, $\l[A \r]$, $\l[A_2 \r]$, $\l[I \r]$, $\l[AI \r]$, and $G$ are the concentrations of the activator mRNA, inhibitor mRNA, monomeric activators, homodimeric activators, inhibitors, and heterodimers. $G$ is the promoter state variable, $G_{\max}=3$ is the total number of binding sites, and $V$ is analogous to the system size. The association and dissociation rates are multiplied by $G$ and $(G_{\rm max}-G)$ because there are multiple combinations of promoters with same number of bound activators due to distributive binding. We reduced the PDMP into a linear PDMP (Appendix E). The Monte Carlo kinetic simulation of the linear PDMP gives similar results to the Full CME of the KB model; see Fig.~\ref{fig:KB}(a-b). In both cases, the stochastic cycles exhibit a well-defined distribution of periods with reduced CV, as previously observed.

\begin{figure*}[t]
\begin{center}
\includegraphics[width=0.98\textwidth]{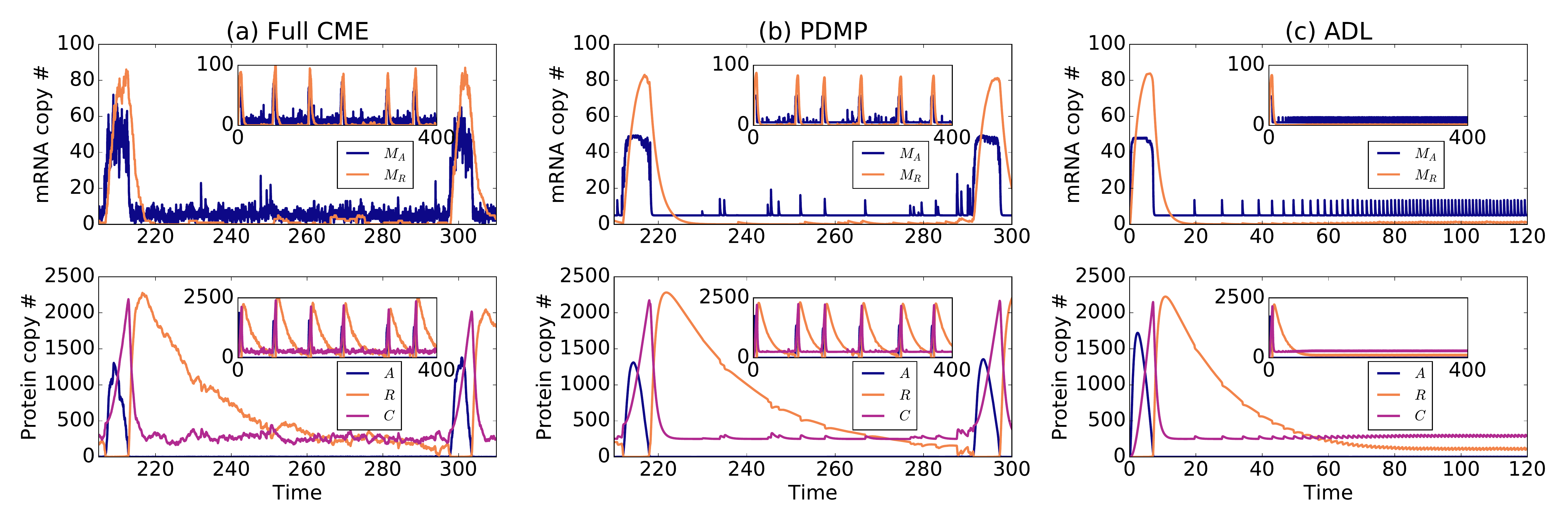}
\end{center}
\caption{
Slow fluctuations in the random binding and unbinding events induce the excitable mode of VKBL model~\cite{vilar2002mechanisms}. Column (a) is the full CME, (b) is the PDMP, and (c) is the ADL of the VKBL model. The first row are the dynamics of the mRNAs and the second are the dynamics of the activator (A), inhibitor (R), and heterodimer complex (C). Insets are longer time series.
}
\label{fig:VKBL}
\end{figure*}

The schematic of the linear PDMP in Fig.~\ref{fig:KBATCPDMP} suggests that the period and coherence improved because the mRNA dynamics introduce a time lag between the change in mRNA production rate and the resulting protein synthesis rate. Thus, even though the transcription rate changes instantaneously upon crossing the boundary when $G = 0 \rightarrow 1$, the mRNA levels will respond and reach a new state on the timescale set by the mRNA degradation rate $\delta_m$.  This lag delays the process of deterministic titration, which requires new inhibitor synthesis, such that $G$ can reach saturation before $x$ is titrated. As a consequence, the KB model now goes through the largest cycle from $G=0$ to $G=3$. Given the importance of the lag, we expect the coherence of stochastic cycling to decrease upon increasing the rate of mRNA degradation and, thus, making the mRNA more responsive to changes in transcription. We tested this idea by rescaling the mRNA degradation $\delta_m = \varpi \bar{\delta}_m$ and protein translation $\beta = \varpi \bar{\beta}$, such that the total protein levels stayed fixed, but mRNA degradation rate could be varied through $\varpi$. Our results in Fig.~\ref{fig:KB}(c) confirm that increasing the mRNA degradation rate via larger $\varpi$ created shorter and less coherent stochastic `mini-cycles’, similar to the idealised ATC which had a measured CV$=0.622$.

We noticed that the variance of the waiting times of binding events in the linear PDMP was much less than those of unbinding events. This motivated us to keep only the stochastic unbinding events whose waiting times are all exponentially distributed and take a deterministic waiting time for the binding events by evaluating the first moment of the cumulative distribution \eqref{eq:survivalYay}. The resulting model is summarized in Fig.~\ref{fig:KBreduced} and is referred to as the reduced PDMP. In the reduced model, the only stochasticity---the random unbinding events---results in a random duration in a series of promoter states which actively produce the inhibitor $I$ (top row of Fig.~\ref{fig:KBreduced}). The excellent agreement between the reduced PDMP and full CME of the KB model suggests that the variability in stochastic cycle times is mostly determined by the stochasticity of unbinding events; see Fig.~\ref{fig:KB}(d). Although the serial nature of unbinding events helps reduce the overall CV and improve the coherence of stochastic cycling, the randomness of unbinding events propagates nonlinearly and contributes to the overall uncertainty of the stochastic cycle period. For example, in each `episode' of serial stochastic unbinding, the number of synthesized inhibitors will be a random quantity that subsequently determines the time to titrate the produced $I$ back to zero (downward arrow) before the promoter state can deterministically cycle back to the actively producing $I$ state ($G=G_{\max}$ and $a=0$).

\subsection{Noise-induced oscillation in the VKBL model}\label{sec:VKBL}
We then turned our attention to the ATC model studied by Vilar et al.~ \cite{vilar2002mechanisms}, whose schematic is shown in Fig.~\ref{fig:complexModels}(b). In the VKBL model, the activator activates itself in addition to the inhibitor and, thus, can exhibit deterministic limit cycles. However, the authors deliberately studied the VKBL model for a parameter set where there were no deterministic limit cycles but the system exhibited excitation-relaxation or noise-induced oscillations. Below, we will use PDMP analysis to show that stochastic promoter fluctuations are responsible for kicking the stable fixed point into an excitable excursion. The VKBL model is given by: 

\al{
\frac{\dd \l[M_A\r]}{\dd t} {}&= \mathbf{1}_{\l\{G_\text{A}=0\r\}} \alpha_A + \mathbf{1}_{\l\{G_\text{A}>0\r\}} \alpha_A'-\delta_{M} \l[M_A\r], \nonumber\\
\frac{\dd \l[M_R\r]}{\dd t} {}&= \mathbf{1}_{\l\{G_\text{R}=0\r\}} \alpha_R + \mathbf{1}_{\l\{G_\text{R}>0\r\}} \alpha_R'-\delta_{M} \l[M_R\r],\nonumber\\
\frac{\dd \l[A\r]}{\dd t} {}&= \beta_A \l[M_A\r] - \delta_A \l[A\r] - \gamma_C \l[A\r] \l[R\r] ,\nonumber\\
\frac{\dd \l[C\r]}{\dd t}{}&= \gamma_C \l[A\r] \l[R\r]  - \delta_A \l[C\r] \\
\frac{\dd \l[R\r]}{\dd t} {}&= \beta_R \l[M_R\r] - \delta_R \l[R\r] - \gamma_C \l[A\r] \l[R\r] + \delta_A \l[C\r],\nonumber\\
G_A=0 {}&\xrightleftharpoons[\theta_A]{\gamma_A \l[A\r]} G_A =1  \nonumber\\
G_R=0 {}&\xrightleftharpoons[\theta_R]{\gamma_R \l[A\r]} G_R = 1. \nonumber
}{eq:VKBLPDMP} 
We adopt the same symbols and parameters of Fig.~5 from the original work \cite{vilar2002mechanisms}, except for discrete $G_\text{A}\in \l\{0,1\r\}$ and $G_\text{R} \in \l\{0,1\r\}$, which represent the number of bound activators on the promoters of A or R.

The sample path of the PDMP faithfully captures the signature of the dynamics of the full CME in the parameter regime with noise-induced oscillations; compare Figs.~\ref{fig:VKBL}(a) and (b). The PDMP only takes into account the stochasticity of the binding and unbinding events (i.e., $G_\text{Z}=0 \leftrightharpoons 1$); the rest of the processes are described by deterministic evolutionary equations. Thus, we can conclude that the noise-induced oscillations in the full CME are due to the discrete binding and unbinding events at the promoter site. In both the full CME and PDMP, the system constantly switches back-and-forth between $G_\text{A}=0 \leftrightharpoons 1$ and produces a bursty activator mRNA population (Fig.~\ref{fig:VKBL}(a-b)). However, occasionally, an unbinding event takes longer than usual which leads to a larger-than-average number of activator mRNAs. This larger-than-average number of activators titrates all the inhibitors ($R$) and the critical accumulation of activator excites the system through a large excursion in the phase space; see Fig.~\ref{fig:VKBL}(b). The alternative deterministic limit of the VKBL model does not exhibit any excitable excursions, as shown in Fig.~\ref{fig:VKBL}(c). By definition, the ADL does not exhibit any variability in the binding and unbinding events. The lack of excitable excursions in the ADL is consistent with the idea that rare fluctuations in the unbinding times are the critical ingredient for generating enough activators ($A$) to titrate all the inhibitors ($R$) in the system.

\section{Discussion and future outlook}\label{sec:discussion}

Dynamical models of gene expression often assume that switching between promoter states (e.g. binding and unbinding of regulatory proteins) takes place at a much shorter timescale than any other processes in the model. This idealisation, known as the quasi steady-state or adiabatic approximation \cite{hornos2005self,ackers1982quantitative}, uses an effective rate inferred from the quasi-stationary distribution of the promoter states.  When the timescale of promoter state switching is comparable to other processes, which is typically the case for natural systems, this approach fails to describe the resulting dynamics accurately. 

In this article, we investigated the stochastic dynamics of biological clocks in the non-adiabatic regime.  Previous work \cite{franccois2005core, karapetyan2015role} demonstrated that time delays, which arise from slow promoter switching in the non-adiabatic regime, are important for the emergence of deterministic limit cycles. These studies modeled the transcription rate as an ensemble-averaged transcription rate of the discrete promoter states. Such a treatment would be precise if one had a large copy number of independent promoters, e.g. models in \cite{wang2016molecular}. However, the copy number of genomic DNA is small in many biological systems and the transcription rate at any given time can only be one of the two discrete values $\beta^b_\text{Z}$ and $\beta^f_\text{Z}$. When the switching timescale is fast (i.e., adiabatic), the promoter state goes through a large numbers of cycles between consecutive transcription events and the effective rate of transcription converges to the ensemble-averaged transcription rate. However, when the switching timescale is slow (i.e., non-adiabatic), the averaged transcription rate cannot capture the nature of alternating rates of transcription events.

The effects of non-adiabatic promoter fluctuations have been investigated, mostly numerically, in the literature of biological clocks. Both the studies of Potoyan and Wolynes \cite{potoyan2014dephasing} and Gonze et al.~\cite{gonze2004emergence} reported that slower switching rates compromise the coherent oscillation seen in the deterministic limit. In a slightly different model, Feng et al.~\cite{feng2012landscape} observed coherent oscillation when the binding and unbinding events were either very fast (adiabatic) or very slow (non-adiabatic). Last, stochastic resonance was reported by Li and Li \cite{li2008noise}, who showed that there exists a `sweet spot' where the promoter switching is neither fast or slow. In the above-mentioned studies, the adopted methods range from direct computation of the eigenvalues of the truncated Chemical Master Equation \cite{potoyan2014dephasing}, direct computation of the stationary distribution \cite{feng2012landscape}, and direct continuous-time Markov simulations and power spectral analyses of the generated sample paths \cite{gonze2004emergence,li2008noise}. Although it is straightforward to carry out these analyses, they reveal little about the mechanisms of stochastic oscillations. For example, these methodologies could not answer why a system with more promoter binding sites exhibits more coherent oscillation, or quantify the impact of mRNA or post-translational reactions, e.g. dimerisation.

We present a mathematical framework to analyse the stochastic dynamics of gene expression in the non-adiabatic regime. In this framework, we begin with the most detailed description of the individual-molecular-based and stochastic dynamics, the chemical master equation, and systematically construct the piecewise deterministic Markov processes (PDMP), which retains the discrete and stochastic switching nature of the genetic states. 
This framework is a natural generalisation of our previous work \cite{hufton2016intrinsic,lin2016gene,lin2016bursting}, and the derived PDMP has been shown to be a powerful mathematical tool to model coloured noise in stochastic gene expression \cite{jia2017simplification,friedman2006linking,lipniacki2006transcriptional,ge2015stochastic,li2016levy,jia2017emergent,bokes2017gene,wang2017likelihood,bokes2017high,2017arXiv171008542L}.
Our analyses showed that for the models we investigated, the PDMP faithfully captures dynamical features of the individual-molecular-based models. We further proposed a scheme to construct an alternative `deterministic description' of the dynamics without invoking the adiabatic assumption. These analytical tools revealed the emergent non-equilibrium transitions between the discrete genetic states in the non-adiabatic regime. In the idealised models, both the ATC and RTC exhibited stochastic cycling in the discrete genetic states.  We showed that a more robust and coherent oscillation (the full cycle) occurred in a regime of slower dissociation rate (small $\theta_\text{Z}$ in the idealised model). The analysis also revealed the interactions between the TF population and the transition between the discrete genetic states, showing that it is necessary to consider the joint process describing the TF dynamics and gene switching dynamics. While the joint process (PDMP) is Markovian, it is known that the TF dynamics alone \cite{bena2006dichotomous} is non-Markovian. In this work, we showed that the gene switching dynamics alone is also non-Markovian.

To illustrate the practicality of these analytical tools, we analysed more sophisticated and detailed models.  Interestingly, in the two models we investigated, the analysis revealed different mechanisms of to induce oscillations. In the KB model \cite{karapetyan2015role}, we found that the oscillation was induced by the slow-transitions between the discrete-genetic states, similar to the idealised models. Nevertheless, in contrast to the idealised model which exhibits similar dynamics when we changed the number of promoter sites $\mathcal{N}_\text{Z}$, the inclusion of the mRNA in the KB model pushes the system to transit through more genetic states. This is because the product of the gene, i.e. mRNA, no longer directly (and abruptly) regulates its own production rate and there is a delay. Since the predominant stochasticity of the system arises from the random unbinding events, the ability to travel through more internal stages decreases the coefficient of variance.  Consequently, more coherent oscillations were observed in the KB model, compared to the idealised models. By applying the analytical tools to the VKBL model \cite{vilar2002mechanisms}, we were able to show that the average (deterministic) genetic switching events were not sufficient to induce the oscillation. Instead, longer-than-average binding events were responsible for exciting the FitzHugh-Nagumo-like system to go through a large excursion in the phase space. Biologically, these longer-than-average binding events are called transcriptional bursting noise. It is straightforward to show that by increasing translational bursting, achieved by simultaneously scaling up the translation rate and scaling down the transcription rate, one can also induce similar oscillations (data not shown).

On a final note, the PDMP is derived from the detailed Chemical Master Equation and can be viewed as a hybrid model which combines the continuous and deterministic TF dynamics and the discrete and stochastic promoter switching dynamics. The PDMP is therefore a promising `bridge model' connecting detailed and mechanistic computational models and highly-idealised discrete-state oscillators \cite{escaff2014globally,rosas2016globally,escaff2014arrays} and phase oscillators \cite{schwabedal2010effective,schwabedal2010effective2,schwabedal2013phase,newby2014effects} which were previously proposed \emph{ad hoc}.

\section{Authors' contribution}
YTL and NEB equally contributed to the study.

\section{Competing interests}
The authors declare no competing interests.

\section{Funding}
YTL was supported by the Center for Nonlinear Studies, Los Alamos National Laboratory and partially by the Engineering and Physical Sciences Research Council EPSRC (UK) (grant no. EP/K037145/1). NEB was supported by the National Institutes of Health Directors New Innovator Award DP2 OD008654-01 and the Burroughs Wellcome Fund CASI Award BWF 1005769.01.

\bibliographystyle{vancouver}
\bibliography{ref}

\begin{appendix}
\begin{table*}[t]
\footnotesize 
\centering
 	\begin{tabular}{  c l c  c c}
    \hline
    Parameter & \multicolumn{1}{c}{Description} & ATC & RTC & Order of reaction\\
	\hline
	\hline 
	$\Omega $ & Characteristic system size & $10^3$ & $10^3$ & N/A \\
	$\lambda $ & Scaling factor of the binding rate $\kappa_\text{Z}$ and unbinding rate $\theta_\text{Z}$, $\text{Z}\in\l\{X,Y\r\}$ & $(1,10^3)$ & $(1,10^3)$ & N/A \\
	$\mathcal{N}_\text{X}$ & Number of binding sites on gene X & 0 & $(1,3)$ &N/A \\
	$\mathcal{N}_\text{Y}$ & Number of binding sites on gene Y & $(1,3)$  & 0 &N/A \\
	$\beta_\text{X}^f$ & Basal production rate of gene X when the number of bound $X <\mathcal{N}_\text{X}$  & 2 &  10& Zeroth order \\
	$\beta_\text{X}^b$ & Activated production rate of gene X when the number of bound $X =\mathcal{N}_\text{X}$  & 2 &  0&Zeroth order \\
	$\beta_\text{Y}^f$ & Basal production rate of gene Y when the number of bound $X <\mathcal{N}_\text{Y}$  & 0 & 2 & Zeroth order \\
	$\beta_\text{Y}^b$ & Activated production rate of gene Y when the number of bound $X =\mathcal{N}_\text{Y}$  & 10 & 2 & Zeroth order \\
	$\delta_X $ &Degradation rate of TF $X$ & 1 & 1 &First order \\
	$\delta_Y $ & Degradation rate of TF $Y$ & 1 & 1 & First order \\
	$\kappa_\text{X}$ & The binding rate of TF $X$ to an empty target promoter site on gene X & $ 0 $ & $0.2 \lambda$ & Second order \\
	$\kappa_\text{Y}$ & The binding rate of TF $X$ to an empty target promoter site on gene Y & $\lambda$ & 0 &Second order \\
	$\theta_\text{X}$ & The dissociation rate of a TF $X$ bound to promoter sites of X & $0$  & $0.4 \lambda$ & First order \\
	$\theta_\text{Y}$ & The dissociation rate of a TF $X$ bound to promoter sites of Y & $0.5\lambda$ & $0$ & First order \\	
	$\alpha$ & The production rate of heterodimer $XY$ & 10 & 10 & Second order \\
	\hline
	\end{tabular}
    \caption{Descriptions and values of the idealised model parameters. 
    }
	\label{table:ATCRTC} 
\end{table*}

\begin{figure}[!t]
\begin{center}
\includegraphics[width=0.44\textwidth]{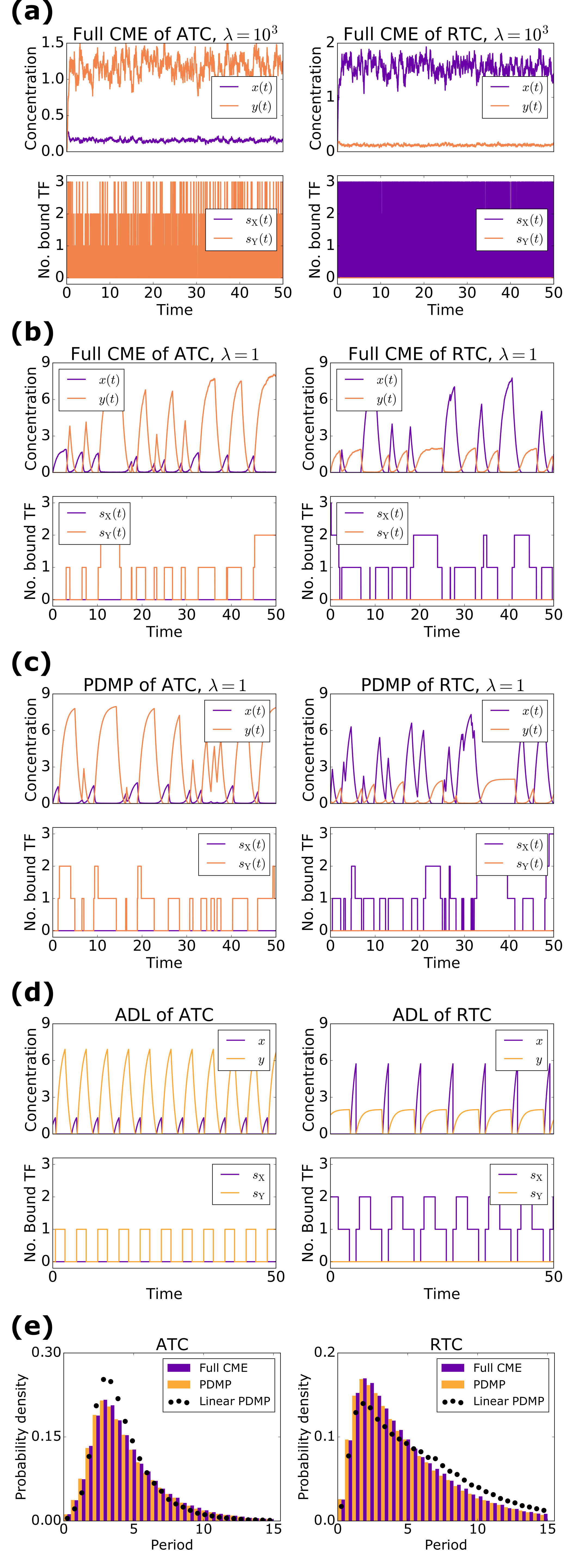}
\end{center}
\caption{
Sample paths of the Full CME model of the ATC and RTC in the (a) adiabatic regime ($\lambda=1000$) and (b) non-adiabatic regime ($\lambda=1$) for multiple binding sites ($\cal{N}_\text{Z}$=3). (c) Sample paths of the constructed piecewise deterministic Markov process when $\lambda=1$ (Sec.~\ref{sec:PDMP}). (d) The alternative deterministic limit of the processes (Sec.~\ref{sec:dATCRTC}). (e) Quantification of the periods of the stochastic cycles. 
}
\label{fig:ATCRTC_N=3}
\end{figure}

\section{Scaling relationship between parameters in the mass action kinetics and parameters in the Full CME model}\label{app:scaling}
The scaling of parameters depends on the order of the reactions. The mapping from the defined mass action rates to the rates in the Full CME simulations are:
\subeq{
\text{0$^\text{th}$ order reactions:}{}& \quad \beta_\text{Z}^f\rightarrow \Omega \beta_\text{Z}^f,\, \beta_\text{Z}^b\rightarrow \Omega \beta_\text{Z}^b, \\
\text{1$^\text{st}$ order reactions:}{}&  \quad \delta_Z\rightarrow \delta_Z,\, \theta_\text{Z} \rightarrow \theta_\text{Z} \\
\text{2$^\text{nd}$ order reactions:}{}&\quad  \alpha \rightarrow \alpha/\Omega, \kappa_\text{Z}\rightarrow \kappa_\text{Z}/\Omega,
}{eq:scaling}
with $\text{Z}\in \l\{\text{X},\text{Y}\r\}$. 

\section{Measuring the periods of stochastic cycles}\label{app:periods}
In the marginal space describing the genetic state ($s_\text{Z}$), the regulatory protein dynamic of $s_\text{Z}=0$ is significantly different from the other states ($s_\text{Z}>0$). This is because we defined the transcription rate of the regulated gene Z to be $\mathbf{1}_{\l\{s_\text{Z}=>0\r\}} \beta^b_\text{Z} + \mathbf{1}_{\l\{s_\text{Z}=0\r\}} \beta^f_\text{Z}$. 
Therefore, we record the times at which each transition $s_\text{Z}=0\rightarrow 1$ occurs, and we define the elapsed time between two consecutive transitions as the period of the stochastic cycle. We measured $10^5$ periods for each model and compared their probability distributions in Fig.~\ref{fig:ATCRTC}(e). 

\section{A kinetic Monte Carlo scheme to generate sample paths of nonlinear PDMPs} \label{app:algo}
When the deterministic part of the PDMP has an analytical solution, Bokes et al.~prescribed a simple algorithm to generate the exact random switching times \cite{bokes2013transcriptional}. We used the Bokes algorithm for our linearised PDMP, which has an analytical solution for the deterministic part. The two-dimensional deterministic system of the full PDMP is governed by a set of nonlinear equations where the nonlinearity comes from the term describing heterodimer formation, $\alpha x y$. As the exact solution is unknown, we used the following numerical scheme to generate exact sample paths that respect the waiting times before the next switching events. In our idealised model of the ATC or RTC, there is only one gene whose promoter states can switch. Our algorithm below generates sample paths for this specific type of network with one promoter $s_Z$, but it can be generalized to more complex, multiple-switching genes. 

\begin{enumerate}
\item {{\bf Initiation.} Initiate the state variable $(x,y,s_Z)$. Here $(x,y)$ are the population density of the TFs, and $s_Z$ is the discrete promoter state; in ATC, $s_Z=s_\text{Y}$ and in RTC, $s_Z=s_\text{X}$. Initiate a ``time of last switching'' $t_0\leftarrow 0$.}
\item {{\bf Generate dissociation time.} When $s_Z>0$, the bound X can dissociate from the promoter sites. The exponentially distributed waiting time is generated by assigning $T_\text{diss}\leftarrow -\log u /\theta$, with $u\sim \text{Unif}(0,1)$. When $s_Z=0$ assign $T_\text{diss} \leftarrow \infty$. }
\item {{\bf Generate a random number to determine the next binding event.} When $s_Z$ is less than the maximum capacity of promoter sites (ATC: $\cal{N}_\text{Y}$; RTC: $\cal{N}_\text{X}$), it is possible to have a binding event in the future. Generate a $u_1 \sim \text{Unif}(0,1)$ for future use. When $s$ is equal to the maximum capacity of the promoter sites, we set $u_1 \leftarrow -1$.}
\item {{\bf Forward integrate the system.} Advance the time by a small $dt\ll1$ to forward integrate the ODE's numerically, update the state $(x,y)$. We implement the integrator using Runge--Kutta method. }
\item {{\bf Check if a binding or dissociation event occurs.} If the time $t>T_\text{diss}$, there was a dissociation event that occurred in the past time step. Update the genetic state $s_Z\leftarrow s_Z-1$. On the other hand, the probability that the system has not bind another TF molecule is
\eq{
\mathbb{P} \l\{T_\text{binding} > t\r\} = \exp \l[- \int_{t_0}^t  \kappa_\text{Z} x(t') \dd t'\r]
}{eq:steve}
We compute this quantity, noting that this can be summed up for each of the time step $\dd t$ numerically. 
If $\mathbb{P} \l\{T_\text{binding} > t\r\} < u_1$ , we know by the inverse transform sampling that a binding event occurred in the past time step, so assign $s_Z \leftarrow s_Z+1$ accordingly. 
}
\item {{\bf Repeat.} If there was a change in the promoter state $s_Z$, then repeat from 2 and register a new $t_0$; otherwise, repeat from 4, until the end of the simulation.}
\end{enumerate}

We remark that if the dynamics are linear and solvable, one can analytically compute the survival function Eq.~\eqref{eq:steve} and derive an more efficient continuous-time sampling technique \cite{bokes2013transcriptional}. 
In the VKBL model, the above algorithm is generalised to the two genetic states $\l(G_A, G_R\r)$. 
We notice that there are four distinct genetic states $\psi_1:=\l(0,0\r)$, $\psi_2:=\l(1,0\r)$, $\psi_3:=\l(0,1\r)$, and $\psi_4:=\l(1,1\r)$, and the possible transitions are $\psi_1 \leftrightharpoons \psi_2 \leftrightharpoons \psi_4 \leftrightharpoons \psi_3 \leftrightharpoons \psi_1 $. Therefore, when the genetic state is in $\psi_1$, we derive two survival functions and use them to perform inverse sampling which generates a first binding event. Similarly, when the genetic state is in $\psi_4$, two exponentially distributed dissociation times are be sampled to determine the first dissociation event. As for genetic states $\psi_2$ and $\psi_3$, they can either transit to $\psi_1$ (by a dissociating event of the bound TF) or $\psi_4$ (by an binding event between the free promoter and a free TF). The random times are sampled similar to the above step 2 to 5.  

\section{PDMP of idealised ATC and RTC}
The PDMP of the ATC is
\al{
\dot{x}={}&\beta_\text{X} - \delta_X x - \alpha x y,\\
\dot{y}={}&\mathbf{1}_{\l\{s_\text{Y}=0\r\}}\beta_\text{Y}^f + \mathbf{1}_{\l\{s_\text{Y}>0\r\}}\beta_\text{Y}^b - \delta_Y y- \alpha x y, \nonumber \\
s_\text{Y} &\xrightarrow{x \kappa_\text{Y}}{} s_\text{Y}+1, \text{ if } 0\le s_\text{Y} < \mathcal{N}_\text{Y},  \nonumber \\
s_\text{Y} &\xrightarrow{\theta_\text{Y}}{} s_\text{Y}-1, \text{ if } 0 < s_\text{Y}  \le \mathcal{N}_\text{Y}. \nonumber
}{}
and the PDMP of the RTC is
\al{
\dot{y}={}& \beta_\text{Y} - \delta_Y y - \alpha x y,\\
\dot{x}={}& \mathbf{1}_{\l\{s_\text{X}=0\r\}}\beta_\text{X}^f + \mathbf{1}_{\l\{s_\text{X}>0\r\}}\beta_\text{X}^b - \delta_X x- \alpha x y, \nonumber \\
s_\text{X} &\xrightarrow{x \kappa_\text{X}}{} s_\text{X}+1, \text{ if } 0\le s_\text{X} < \mathcal{N}_\text{X},  \nonumber \\
s_\text{X} &\xrightarrow{\theta_\text{X}}{} s_\text{X}-1, \text{ if } 0 < s_\text{X}  \le \mathcal{N}_\text{X}. \nonumber
}{}
where $\mathcal{N}_\text{Z}$ is the number of promoter states. 

\section{Linear PDMP of the KB model} 
We performed the following model reduction for the nonlinear PDMP in Eq. ~\eqref{eq:KBPDMP} by imposing the following assumptions based on the parameters used in \cite{karapetyan2015role}:\\

\noindent{\bf Irreversible heterodimersation.} The heterodimerisation is much larger than the reverse rate, and we approximate it as an irreversible process. Thus, $\l[AI \r]$ is ignored.\\

\noindent{\bf Fast homodimerisation and dissociation.} The homodimerisation and dissociation occurs at a much faster timescale than other processes, so the concentrations of $\l[A \r]$ and $\l[A_2 \r]$ satisfy the quasi-stationary approximation---$\gamma \l[A \r]^2 \approx \epsilon_1 \l[A_2 \r]$---at any given time. For simplicity, we define $\l[\mathcal{A}\r]:=\l[A \r] + 2\l[A_2 \r](t)$ as the total number of activators in monomeric and dimeric form.\\

\noindent{\bf Linearisation.} We assume that at any given time, either activators or inhibitors are dominant such that the other becomes a limiting factor. This is the approximation we proposed in Sec.~\ref{sec:dATCRTC} for the idealized model. After this linearisation, the dynamics of $\l[{r}_A \r]$, $\l[{r}_I \r]$, $\l[\mathcal{A} \r]$, and $\l[I \r]$ are all analytically tractable. 
In the long run, $\l[{r}_A \r] \rightarrow \rho_0/\delta_m$, and $\l[{r}_I \r]$ relax exponentially to the fixed points $\rho_f/\delta_m$ and $\rho_b/\delta_m$ when $G=0$ and $G>0$ respectively.
The activation rate of the linearised PDMP is proportional to the concentration of the dimeric activators $A_2$, which can be solved by equating
\eq{
\l[\mathcal{A}\r] = \l[A \r]+2 \l[A_2 \r] = \sqrt{\frac{\epsilon_1}{\gamma} \l[A_2\r]} + 2 \l[A_2\r],
}{}
using the adiabatic approximation $\gamma \l[A \r]^2 = \epsilon_1 \l[A_2 \r]$ in the last step.
The survival function of the waiting time of the next binding event can be formulated as follows: let the random time to the next binding event to be $\tau$, 
\eq{
\pr{\tau > t \vert G(t_0) = G_0 } = e^{ - \alpha \l(G_{\max} - G_0 \r)\int_{t_0}^{t} \l[A_2\r]\l(t'\r) \dd t' }.
}{eq:survivalYay}
The survival function is then used to generate the stochastic waiting time to next binding event using the inverse transform sampling method.

\end{appendix}
\end{document}